\documentclass[aps,prd,twocolumn,a4paper,superscriptaddress,nofootinbib]{revtex4-1}
\usepackage{amsmath}
\usepackage{graphicx}
\usepackage{multirow}
\usepackage{latexsym}
\usepackage[british]{babel} 
\usepackage{soul}
\usepackage[colorlinks,linkcolor=blue,citecolor=blue,urlcolor=blue ]{hyperref}
\begin{document}

\newcommand{\be}{\begin{equation}}
\newcommand{\ee}{\end{equation}}
\newcommand{\bq}{\begin{eqnarray}}
\newcommand{\eq}{\end{eqnarray}}
\newcommand{\bsq}{\begin{subequations}}
\newcommand{\esq}{\end{subequations}}
\newcommand{\bc}{\begin{center}}
\newcommand{\ec}{\end{center}}
\newcommand{\lan}{\langle}
\newcommand{\ran}{\rangle}
\newcommand{\eps}{\epsilon}
\newcommand{\al}{\alpha}

\title{Varying Alpha Generalized Dirac-Born-Infeld Models}
\author{V. C. Tavares}
\email{up201705797@fc.up.pt}
\affiliation{Centro de Astrof\'{\i}sica, Universidade do Porto, Rua das Estrelas, 4150-762 Porto, Portugal}
\affiliation{Faculdade de Ci\^encias, Universidade do Porto, Rua do Campo Alegre 687, 4169-007 Porto, Portugal}
\author{C. J. A. P. Martins}
\email{Carlos.Martins@astro.up.pt}
\affiliation{Centro de Astrof\'{\i}sica, Universidade do Porto, Rua das Estrelas, 4150-762 Porto, Portugal}
\affiliation{Instituto de Astrof\'{\i}sica e Ci\^encias do Espa\c co, Universidade do Porto, Rua das Estrelas, 4150-762 Porto, Portugal}
\date{14 November 2020}
\begin{abstract}
We study the cosmological consequences of a class of Dirac-Born-Infeld models, and assess their viability as a candidate for the recent acceleration of the Universe. The model includes both the rolling tachyon field and the generalized Chaplygin gas models as particular limits, and phenomenologically each of these provides a possible mechanism for a deviation of the value of the dark energy equation of state from its canonical (cosmological constant) value. The field-dependent potential that is characteristic of the rolling tachyon also leads to variations of the fine-structure constant $\alpha$, implying that the model can be constrained both by standard cosmological probes and by astrophysical measurements of $\alpha$. Our analysis, using the latest available low-redshfit data and local constraints from atomic clock and weak equivalence principle experiments, shows that the two possible deviations of the dark energy equation of state are constrained to be $\log_{10}{(1+w_0)_V}<-7.85$ and $\log_{10}{(1+w_0)_C}<-0.85$, respectively for the rolling tachyon and Chaplygin components, both being at the $95.4\%$ confidence level (although the latter depends on the choice of priors, in a way that we quantify). Alternatively, the $95.4\%$ confidence level bound on the dimensionless slope of the potential is $\log_{10}{\lambda}<-5.36$. This confirms previous analyses indicating that in these models the potential needs to be extremely flat.
\end{abstract}

\maketitle

\section{\label{intro}Introduction}

A key goal of contemporary cosmology is the search for the identification and characterization of the mechanism responsible for the observed acceleration of the universe. In a nutshell, the main possibilities are: a cosmological constant (which has a minimal number of additional parameters but a value requiring fine-tuning), modifications of the behaviour of gravity (for which there arguably is no compelling physical evidence), or additional dynamical degrees of freedom (particularly scalar fields, which are known to be among Nature's building blocks) \cite{Copeland,Frieman,Joyce}.

The first of these scenarios is, broadly speaking, in agreement with the currently available data, but nevertheless there are several observational hints of inconsistencies. While individually none of these seems to be particularly compelling (or free from caveats), when they are taken together they certainly provide a robust motivation for exploring theoretical alternatives and testing them against the available data \cite{Huterer}.

A recent work of our team provided low-redshift cosmological constraints on rolling tachyon dark energy models \cite{Roll}. These are a class of Dirac-Born-Infeld (hereafter DBI) dark energy models, which are well motivated in string theory \cite{Sen1,Sen2}. From the observational point of view, one of the interesting features of these models is a coupling between gauge fields and the scalar field responsible for the universe’s acceleration \cite{Garousi}. This implies that in these models the field dynamics naturally leads to a time variation of the fine-structure constant, $\alpha$, which can be constrained with astrophysical observations as well as local tests using atomic clocks and weak equivalence principle experiments \cite{ROPP}.

In this work we continue the aforementioned previous study, specifically by introducing and exploring the consequences of an extended phenomenological class of generalized DBI models, whose limits include both the rolling tachyon scenario discussed in \cite{Roll} and also the generalized Chaplygin gas \cite{Bento,Beca}. This extension is phenomenologically interesting because the deviations of the dark energy equation of state from the cosmological constant can be decomposed into two components, one of which corresponds to the rolling tachyon limit while the other corresponds to the Chaplygin limit.

On the other hand, an interesting difference arises when it comes to the variation of $\alpha$, since only the rolling tachyon part contributes to it. This implies that its part of the field dynamics (and the corresponding equation of state) will be significantly more constrained than that of the Chaplygin component. The model thus serves to illustrate the relative constraining power of background low-redshift astrophysical and cosmological observables on the dynamics of non-canonical DBI-type scalar fields.

The plan of the paper is as follows. In Sect. \ref{methods} we start by introducing the general formalism used in this work, before reviewing, for the sake of completeness, its application to the canonical, rolling tachyon and generalized Chaplygin gas cases. The generalized class of DBI models is then presented in Sect. \ref{dbi}, where we specifically discuss their low redshift behaviour, which is the focus of our work. In Sect. \ref{results} we use a combination of low-redshift background cosmology data and astrophysical and atomic clock measurements of $\alpha$ to constrain these models. We discuss the impact of adding the astrophysical measurements to the cosmological data, and by considering two alternative parametrizations we also quantify the sensitivity of our results on the choice of priors (as well as on the choice of parametrization itself). Finally, we present our conclusions in Sect. \ref{outlook}.

\section{\label{methods}Formalism}

Consider the generic action for a classical scalar field, denoted $\phi$,
\be
S=\int d^4x\sqrt{-g}{\cal L}(X,\phi)\,,
\ee
where
\be
X=-\frac{1}{2}\nabla^\mu\phi\nabla_\mu\phi\,.
\ee
The proper pressure and energy density of the scalar field (which reduce to the usual pressure and density in the field's rest frame where the momentum density vanishes), are
\be
p={\cal L}
\ee
\be
\epsilon=2Xp_{,X}-p\,,
\ee
with the energy-momentum tensor being
\be
T_{\mu\nu}=pg_{\mu\nu}+p_{,X}\nabla_\mu\phi\nabla_\nu\phi\,.
\ee
The field equation of state is
\be
w=\frac{p}{\epsilon}=\frac{p}{2Xp_{,X}-p}\,.
\ee
from which we see that $w=-1$ (vacuum energy) requires $2Xp_{,X}=0$, while the opposite limit $w=1$ (known as kination) requires $p=Xp_{,X}$ (that is, $p\propto X$). Finally, the sound speed is
\be
c_s^2=\frac{p_{,X}}{\epsilon_{,X}}=\frac{p_{,X}}{2Xp_{,XX}+p_{,X}}\,,
\ee
from which we see that $c_s^2=1$ requires $2Xp_{,XX}=0$, and we can write
\be
w_{,X}=(c_s^2-w)\frac{\epsilon_{,X}}{\epsilon}\,.
\ee
Finally, it is also useful to define the slow-roll parameter
\begin{equation}
s = - \frac{3 X}{p} \frac{\partial p}{\partial X}\,.
\end{equation}

For the case of a perfect fluid we can write
\be
{\dot \epsilon}+3H(\epsilon+p)=0
\ee
or equivalently
\be
{\dot\epsilon}+6HXp_{,X}=0\,,
\ee
which can be expanded and written as follows
\be\label{generic1}
2X{\dot X}p_{,XX}+(2Xp_{,X\phi}-p_{,\phi}){\dot\phi}+({\dot X}+6HX)p_{,X}=0
\ee
or equivalently
\be\label{generic2}
({\dot\phi}^2p_{,XX}+p_{,X}){\ddot\phi}+3Hp_{,X}{\dot\phi}+({\dot\phi}^2p_{,X\phi}-p_{,\phi})=0\,.
\ee
These will be applied to the various models being studied.

In our cosmological setting we will be considering flat Friedmann-Lema\'{\i}tre-Robertson-Walker models. The previously defined energy density $\epsilon$ and the mass density $\rho$ are generically related via $\epsilon=\rho c^2$; further choosing units with $c=1$ the two can be used interchangeably, and in what follows we will express relevant quantities in terms of $\rho$. Thus the Einstein and continuity equations can be written
\be
H^2=\frac{\kappa^2}{3}\rho
\ee
\be
{\dot H}=-\frac{1}{2}\kappa^2(\rho+p)
\ee
\be
{\dot\rho}=-3H(\rho+p)\,,
\ee
where for convenience we have also defined $\kappa^2=8\pi G$. We can also write the overall equation of state
\be
1+w=-\frac{2}{3}\frac{\dot H}{H^2}\,.
\ee
Since in this work we will be concerned with the recent dynamics of the universe we will be assuming a universe with two components, matter (in principle both baryonic and dark) and a dark energy fluid with an equation of state $\rho=\rho_m+\rho_\phi$ which can be explicitly written as
\be\label{defeos}
1+w_\phi=-\frac{1}{3}\frac{d\ln{\rho_\phi}}{d\ln{a}}=\frac{1}{3}\frac{d\ln{\rho_\phi}}{d\ln{(1+z)}}\,.
\ee
This can be constrained through the usual cosmological probes; in our case we will use Type Ia supernova data and Hubble parameter measurements, to be further discussed in Sect. \ref{results}.

Our second observational probe of these models will be the value of the fine-structure constant $\alpha$, whose possible variation stems from the coupling of a putative new degree of freedom (such as a scalar field) to the electromagnetic sector of the Lagrangian
\be
{\cal L}_F=-\frac{1}{4}B_F(\phi)F_{\mu\nu}F^{\mu\nu}\,,
\ee
where the gauge kinetic function can be Taylor-expanded \cite{Carroll,Dvali,Chiba}
\be
B_F(\phi)=1-\zeta\kappa(\phi-\phi_0)
\ee
and therefore the relative variation (using the present-day value as reference) is given by
\be
\frac{\Delta\alpha}{\alpha}=\frac{\alpha-\alpha_0}{\alpha_0}=B_F^{-1}-1=\zeta\kappa(\phi-\phi_0)\,.
\ee
Astrophysical measurements of $\alpha$ (to be further discussed in Sect. \ref{results}) constrain its value at various non-zero redshifts, while its current drift rate is constrained by laboratory measurements with atomic clocks.

Moreover, since in these models the new degree of freedom inevitably couples to nucleons (through the $\alpha$ dependence of their masses), it also leads to to violations of the Weak Equivalence Principle \cite{Dvali,Chiba}. It follows that the coupling $\zeta$ will be related to the Eotvos parameter $\eta$, for which there are local experimental constraints. In what follows we use the estimate from \cite{Dvali}
\be
\eta=10^{-3}\zeta^2\,,
\ee
though we note that somewhat different estimates exist \cite{Chiba,Damour}.

We now highlight how this applies to representative examples, both to establish connections with previously obtained results and also because they will be the building blocks for our generalized class of models.

\subsection{Canonical scalar field}

The simplest case is that of a canonical quintessence field, for which we have
\be
p=X-V(\phi)\,,
\ee
from which we easily recover the Klein-Gordon equation
\be
{\ddot\phi}+3H{\dot\phi}+V_{,\phi}=0\,.
\ee
The slow-roll parameter is
\be
s=-\frac{3}{2}\frac{{\dot\phi}^2}{X-V}\,,
\ee
confirming that a slowly moving field is slow-rolling. Then from
\be
2Xp_{,X}=(1+w_\phi)\epsilon
\ee
we successively have
\be
1+w_\phi=\frac{{\dot\phi}^2}{\epsilon}=\frac{(\kappa\phi')^2H^2}{\kappa^2\rho_\phi}=\frac{(\kappa\phi')^2(\rho_m+\rho_\phi)}{3\rho_\phi}=\frac{(\kappa\phi')^2}{3\Omega_\phi}\,,
\ee
and assuming that the field is rolling down the potential (thus $\phi'<0$) we end up with the following expression for the evolution of $\alpha$
\be
\frac{\Delta\alpha}{\alpha}=\zeta\int_0^z\sqrt{3\Omega_\phi(z')(1+w_\phi(z'))}\frac{dz'}{1+z'}\,.
\ee
Its present-day drift rate is
\be
\frac{1}{H_0}\left(\frac{\dot\alpha}{\alpha}\right)_0=-\zeta\sqrt{3\Omega_{\phi}(1+w_0)}\,.
\ee
These have been previously discussed, {\it inter alia}, in \cite{Lidsey,Amendola}, and analogous results exist for phantom fields \cite{Pauline}. 

\subsection{Tachyon field}

These models were introduced in \cite{Sen1,Sen2}, and their cosmological consequences have been previously studied in detail in \cite{Roll}. The Lagrangian of the tachyon part of the DBI action can be written
\be
{\cal L_{\rm tac}}=-V(\phi)\sqrt{1-\partial_a\phi\partial^a\phi}\,,
\ee
with the energy density and pressure being given by
\be
\rho_\phi=\frac{V(\phi)}{\sqrt{1-\partial_a\phi\partial^a\phi}}
\ee
\be
p_\phi=-V(\phi)\sqrt{1-\partial_a\phi\partial^a\phi}
\ee
which implies that
\be
p=-\frac{V^2(\phi)}{\rho}\,.
\ee
The dynamical equation for the tachyon field is
\be
\frac{\ddot \phi}{1-{\dot\phi}^2}+3H{\dot\phi}+\frac{1}{V}\frac{dV}{d\phi}=0\,.
\ee
For a homogeneous field, the tachyon field equation of state and sound speed are
\be
w_\phi={\dot\phi}^2-1\ge-1\,,
\ee
\be
c^2_s=1-{\dot\phi}^2\le1\,,
\ee
with the equation of state and density evolving as
\be
{\dot w}_\phi=2w_\phi\left[3H(1+w_\phi)+\frac{1}{V}\frac{dV}{d\phi}{\dot\phi}\right]\,,
\ee
\be\label{conseq0}
{\dot\rho_\phi}=-3H(1+w_\phi)\rho_\phi=-3H\rho_\phi {\dot\phi}^2\,.
\ee

In these models the slow-roll parameter is
\be
s=\frac{3}{2}\frac{{\dot\phi}^2}{1-{\dot\phi}^2}\,,
\ee
and clearly the field is constrained to be slow-rolling, even more so than in canonical models, given  the strong constraints from $\alpha$ variations as shown in \cite{Roll} and further confirmed below. Therefore the scalar field equation can be approximated to
\be
3H{\dot\phi}\propto -\frac{d\ln{V}}{d\phi}\,.
\ee
This leads to the following Friedmann equation
\be
\frac{H^2}{H_0^2}=\Omega_m(1+z)^3+(1-\Omega_m)\left[1+\frac{\lambda^2}{9} f_V(\Omega_m,z)\right]\,,
\ee
where we have defined the dynamically relevant dimensionless parameter (the slope of the potential function)
\be\label{deflambda}
\lambda=\frac{1}{H_0}\left(\frac{V'}{V}\right)_0\,.
\ee
and the redshift-dependent function $f_V(z,\Omega_m)$ is \footnote{ Note that this $f_V$ differs from the analogous function $f$ previously defined in \cite{Roll} by a factor of two. The convenience of this choice will become clearer in what follows.}
\be\label{deffv}
f_V(\Omega_m,z)=\frac{2}{\sqrt{1-\Omega_m}}\ln{\frac{(1+\sqrt{1-\Omega_m})(1+z)^{3/2}}{\sqrt{1-\Omega_m}+E_\Lambda}}\,,
\ee
where for convenience we also defined
\be
E^2_\Lambda(\Omega_m,z)=\Omega_m(1+z)^3+1-\Omega_m\,.
\ee
The dark energy equation of state in these models has the form
\be
1+w_\phi={\dot\phi^2}=\frac{\lambda^2}{9+\lambda^2f_V}\frac{\sqrt{1-\Omega_m}+E_\Lambda}{E^2_\Lambda+\sqrt{(1-\Omega_m)}E_\Lambda}\,,
\ee

In physical terms, the field speed parametrizes the deviation of the dark energy equation of state from the cosmological constant value, and the equation of state $(1+w_\phi)$ tends to zero at high redshifts---in other words, these are thawing dark energy models. Its present-day value is
\be
1+w_0={\dot\phi^2}_0=\frac{\lambda^2}{9}\,.
\ee
Starting with Eq. \ref{conseq0} and using ${\dot\phi}=-\lambda/3$ we can approximately write
\be
\frac{\rho_\phi}{\rho_{crit}}\propto (1+z)^{\lambda^2/3}\sim  (1+z)^{3(1+w_0)}
\ee
which in the low-redshift limit yields
\be
\frac{\rho_\phi}{\rho_{crit}}\propto \left[1+3(1+w_0)z\right]\,.
\ee

The $\alpha$ variation in these models was first considered by \cite{Garousi} who have shown that the fine-structure constant is inversely proportional to the tachyon potential. This was further explored in \cite{Roll}, where it has been shown that the redshift evolution of $\alpha$ can be written
\be
\frac{\Delta\alpha}{\alpha}=-\frac{\lambda^2}{9}f_V(\Omega_m,z)\,,
\ee
implying that in these models the fine-structure constant is always smaller in the past. In this case the present-day rate of change of the fine-structure constant is
\be
\frac{1}{H_0}\left(\frac{\dot\alpha}{\alpha}\right)_0= \frac{1}{3H_0^2}\left(\frac{V'}{V}\right)^2_0=\frac{1}{3}\lambda^2=3{\dot\phi_0}^2=3(1+w_0)\,.
\ee
The work of \cite{Roll} further shows that in these models $w_0$ is effectively indistinguishable from a cosmological constant, although they can have a distinctive astrophysical variation of $\alpha$. This also implies that the field speed today must be tiny
\be
{\dot\phi_0}\leq 10^{-3}\,,
\ee
which further justifies our slow-roll approximation.

\subsection{Generalized Chaplygin gas}

The distinguishing phenomenological feature of the generalized Chaplygin gas is its equation of state\footnote{In most of the previous literature, the exponent in the density is denoted $\alpha$; in what follows we denote it $\beta$ to avoid confusion with the fine-structure constant.}
\be
p=-\frac{A}{\rho^\beta}\,;
\ee
the original Chaplygin gas has $\beta=1$, and in the limit $\beta\longrightarrow0$ we recover $\Lambda$CDM. A non-canonical isentropic perfect fluid description can be obtained from a generalized Born-Infeld-type action \cite{Bento,Beca}, with
\be\label{lagcg}
p(\phi,X)=-A^{1/(1+\beta)}\left[1-(2X)^{(1+\beta)/2\beta}\right]^{\beta/(1+\beta)}\,.
\ee
Since $p=p(\phi,X)$ this is a perfect fluid; furthermore, since $p=p(X)$, it is also an adiabatic fluid.

For the case $\beta=1$ some motivation for the above equation of state can be found in terms of a gas of $(d+2)$ branes in $d$ dimensions, but no similar motivation exists for the case $\beta\neq 1$. In any case, one may think of the generalized Chaplygin gas as a phenomenological toy model with which to study some features that may also apply to other models. The model has been observationally constrained under various different approximations and assumptions \cite{Gorini,Bean,Amendola2,Sandvik,Park,Marttens}.

The original motivation of the generalized Chaplygin gas was as a unified dark energy fluid, simultaneously replacing dark matter and dark energy. Thus the only component apart from it, in the low-redshift universe, should be ordinary baryonic matter. In the background, it behaves as normal matter at early times but as a cosmological constant at sufficient late times. From Eq. \ref{defeos}, one finds that its energy density behaves as
\be\label{chapbase}
\frac{\rho_{Ch}(a)}{\rho_{crit}}= \Omega_{Ch} \left[-w_0+\frac{1+w_0}{a^{3(1+\beta)}}\right]^{1/(1+\beta)}\,;
\ee
it is worthy of note that in the low-redshift limit this behaves as
\be
\frac{\rho_{Ch}}{\rho_{crit}}=\Omega_{Ch}\left[1+3(1+w_0)z\right]\,,
\ee
which matches the result for rolling tachyons in the previous sub-section.

The generalized Chaplygin gas equation of state can be written
\be\label{eoschap}
1+w=\left({\dot\phi^2}\right)^{\frac{1+\beta}{2\beta}}\,,
\ee
or equivalently
\be
1+w=\frac{(1+w_0)(1+z)^{3(1+\beta)}}{(1+w_0)(1+z)^{3(1+\beta)}-w_0}
\ee
which can also be used to re-write
\be
\frac{\rho_{Ch}}{\rho_{crit}}= \Omega_{Ch} \left(\frac{w_0}{w}\right)^{1/(1+\beta)}\,,
\ee
and finally the slow-roll parameter has the following form 
\be\label{slowchap}
s=\frac{3}{2}\frac{\left({\dot\phi}^2\right)^{\frac{1+\beta}{2\beta}}}{1-\left({\dot\phi}^2\right)^{\frac{1+\beta}{2\beta}}}\,;
\ee
both of these coincide with those for the rolling tachyon when one chooses $\beta=1$. The reason why the equations of state $w$ are the same (when expressed in terms of ${\dot\phi}$) is that $p$ has the same dependence on $X$, while the fact that $V$ is constant (or not) is not relevant for this particular purpose.

It is also useful to note that from the equation of state definition we have
\be
A=-w_0(\Omega_{Ch}\rho_{crit})^{1+\beta}\,,
\ee
which fixes the value of the constant parameter $A$. At early times the total (effective) matter density is
\be
\Omega_m=\Omega_b+\Omega_{Ch}(1+w_0)^{1/(1+\beta)}
\ee
and for $\beta\to0$ the background evolution approaches that of the standard $\Lambda$CDM model, with
\be
\Omega_\Lambda=-w_0\Omega_{Ch}
\ee
\be
\Omega_m=\Omega_b+(1+w_0)\Omega_{Ch}\,.
\ee
Moreover, as $w_0\to-1$ the generalized Chaplygin gas evolution approaches that of a cosmological constant, regardless of the value of $\beta$. From the Raychaudhuri equation one also finds that in the background the acceleration starts at
\be
a_\star^{3(1+\beta)}=-\frac{1+w_0}{2w_0}\,;
\ee
in particular this occurs today ($a_\star=1$) for $w_0=-1/3$.

Note that if one takes the standard (physically motivated) approach and sees the generalized Chaplygin gas as the result of a particular class of Born-Infeld-type action with a constant potential, then according to the results of \cite{Sen1,Sen2,Garousi} there will be no variation of $\alpha$ in these models. Some authors have nevertheless considered possible variations of $\alpha$ in these models by making various non-standard assumptions \cite{Feng,Wei,Tayebi}; we will not follow this ad hoc approach here.

\section{\label{dbi}A generalized DBI model}

We will now discuss a generalized DBI-type model, which includes both the rolling tachyon and the generalized Chaplygin gas as particular cases. We have seen in the previous section that the rolling tachyon case corresponds to the equation of state
\be
p=-\frac{V^2(\phi)}{\rho}\,,
\ee
so we can envisage further generalizing Eq. (\ref{lagcg}) by making the constant $A$ depend on the scalar field---in other words, making it the field potential. In this case we will no longer have an adiabatic fluid, but will still have a perfect one. Specifically we need to make the identification
\be
A\longrightarrow V^2(\phi)\,,
\ee
and naturally the constant-potential limit takes us back to the generalized Chaplygin gas limit, while the $\beta=1$ limit takes us back to the rolling tachyon.

Clearly, the field pressure and density now have the form
\be\label{ladbip}
p(\phi,X)=-V(\phi)^{2/(1+\beta)}\left[1-(2X)^{(1+\beta)/2\beta}\right]^{\beta/(1+\beta)}\,.
\ee
\be\label{ladbirho}
\rho(\phi,X)=V(\phi)^{2/(1+\beta)}\left[1-(2X)^{(1+\beta)/2\beta}\right]^{-1/(1+\beta)}\,.
\ee
In this case the slow-roll parameter has exactly the same form as the Chaplygin one, that is Equation (\ref{slowchap}), so the slow-roll approximation will be adequate for any model that aims to account for dark energy and the recent acceleration of the universe.

The background equation of state itself can be written
\be
w=-\frac{V^2(\phi)}{\rho^{1+\beta}}={\dot\phi}^{1+1/\beta}-1=-\xi\,,
\ee
which is again identical to the Chaplygin case, cf. Eq. (\ref{eoschap}). For later convenience we also defined the parameter $\xi$, and as expected we see that the equation of state can evolve from $w\sim0$ at early times to $w\sim-1$ at late times. Differentiating we get
\be
{\dot w}=(1+\beta)w{\dot\phi}^{1/\beta}\left[3H{\dot\phi}+\frac{2}{1+\beta}\frac{{\dot\phi}^2}{1-\xi}\frac{V'}{V}\right]\,.
\ee
while the sound speed has the simple form
\be
c^2_s=\beta\left[1-{\dot\phi}^{1+1/\beta}\right]\,;\label{eq:cs_ch}
\ee
note that since $w$ is always negative, the sound speed is well-behaved for $0\le\beta\le1$. It is also possible to write
\be
\frac{\rho_{Ch}}{\rho_{crit}}=\Omega_{Ch} \left(\frac{w_0V(\phi)^2}{wV(\phi_0)^2}\right)^{1/(1+\beta)}\,.
\ee

Now, for the generalized DBI we can write
\be
p(X)=-\frac{V^2}{\rho^\beta}=-\left[V^2\xi^\beta\right]^{1/1+\beta}
\ee
where
\be
\xi=\frac{V^2}{\rho^{1+\beta}}=1-(2X)^{(1+\beta)/2\beta}=-w\,.
\ee
and therefore the previously introduced generic Eqs. (\ref{generic1}-\ref{generic2}) become
\be
\frac{1-\xi}{\beta\xi(2X)}{\dot X}+3(1-\xi)H+\frac{2}{1+\beta}{\dot\phi}\frac{V'}{V}=0
\ee
or equivalently
\be
\frac{\ddot\phi}{\beta\xi}+3H{\dot\phi}+\frac{2}{1+\beta}\frac{\dot\phi^2}{1-\xi}\frac{V'}{V}=0\,.
\ee
In the limit $V=const.$ we recover the generalized Chaplygin gas case
\be
{\dot X}+6\beta HX\xi=0
\ee
\be
{\ddot \phi}+3\beta H\xi {\dot\phi}=0\,.
\ee
On the other hand, for $\beta=1$ we have the rolling tachyon case
\be
{\dot X}+6HX\xi+\frac{V_{,\phi}}{V}\xi{\dot\phi}=0
\ee
\be
\frac{\ddot \phi}{1-{\dot\phi}^2}+3H{\dot\phi}+\frac{1}{V}\frac{dV}{d\phi}=0\,.
\ee

\subsection{Low-redshift evolution}

For the purpose of our comparison to low-redshft cosmological and astrophysical observations, it is sufficient to consider the slow-roll limit
\be
3H{\dot\phi}^{1/\beta}+\frac{2}{1+\beta}\frac{V'}{V}=0\,.
\ee
Thus a simple generalization of the previous analysis leads to a Friedmann equation
\begin{widetext}
\be
\frac{H^2}{H_0^2}=\Omega_m(1+z)^3+(1-\Omega_m)\left[-w_0+(1+w_0)(1+z)^{3(1+\beta)}\right]^{1/(1+\beta)}\left[1+\left(\frac{2\lambda}{3}\right)^{1+\beta}\frac{f_V(\Omega_m,z)}{2(1+\beta)^{\beta}}\right]\,,
\ee
\end{widetext}
with $\lambda$ and $f_V$ defined as before, cf. Eqs. (\ref{deflambda}) and (\ref{deffv}) respectively. Note that this includes both the Chaplygin and potential corrections to the canonical $\Lambda$CDM model, with the caveat that in this general case $w_0$ is no longer the present-day dark energy equation of state. Instead we can define an effective equation of state with two different contributions, coming from the Chaplygin and potential terms
\be
(1+w_{eff})=(1+w_0)_C + (1+w_0)_V\,,
\ee
where
\be\label{vbetalambda}
(1+w_0)_V=\left(\frac{2}{1+\beta}\right)^\beta \left(\frac{\lambda}{3}\right)^{(1+\beta)}\,,
\ee
which manifestly has the appropriate behaviour in the two limiting cases. We can also re-write the Friedmann equation as
\begin{widetext}
\be
\frac{H^2}{H_0^2}=\Omega_m(1+z)^3+(1-\Omega_m)\left[1+(1+w_0)_Cf_C(\beta,z)\right]^{1/(1+\beta)}\left[1+(1+w_0)_Vf_V(\Omega_m,z)\right]\,,
\ee
where, by analogy with the $f_V(\Omega_m,z)$ term, we have defined
\be
f_C(\beta,z)=(1+z)^{3(1+\beta)}-1\,.
\ee

For the $\alpha$ variation, for the aforementioned reason we will only have the contribution of the changing potential (but not that of the Chaplygin part) and therefore we have an extension of the rolling tachyon behaviour
\be
\frac{\Delta\alpha}{\alpha}\simeq -\left(\frac{V'}{V}\right)_0(\phi-\phi_0)=-\left(\frac{2\lambda}{3}\right)^{1+\beta}\frac{f_V(\Omega_m,z)}{2(1+\beta)^{\beta}}=-(1+w_0)_Vf_V(\Omega_m,z)\,.
\ee
\end{widetext}
Finally, for the present-day drift rate of $\alpha$, which is constrained by atomic clocks, we find
\be\label{lamw0}
\frac{1}{H_0}\left(\frac{\dot\alpha}{\alpha}\right)_0=\left(\frac{2}{3(1+\beta)}\right)^\beta\lambda^{1+\beta}=3(1+w_0)_V\,,
\ee
while for the effective coupling, relevant for the Eotvos parameter constraints, we have
\be
\zeta^2=\left[3(1+w_0)_V\right]^{2/(1+\beta)}\left(\frac{2}{3(1+\beta)}\right)^{-2\beta/(1+\beta)}=\lambda^2\,.
\ee
Again, one can easily confirm that the previous models are recovered in the appropriate limits.

Overall, this parametrization makes it explicit that the $\alpha$ variations are driven by the potential term, while the cosmological evolution is driven by it but also by the Chaplygin part. This is observationally interesting, since the astrophysical measurements of $\alpha$ will strongly constrain $(1+w_0)_V$, leaving the $(1+w_0)_C$ to be constrained by cosmological data.

\section{\label{results}Observational constraints}

We constrain this class of models using a combination of low-redshift background cosmological and astrophysical observations. The cosmology data consists of the Pantheon\footnote{We note that the reliability of this dataset has recently been questioned \cite{Steinhardt}.} catalogue of Type Ia supernovas \cite{Riess} (including its covariance matrix), and a compilation of 38 Hubble parameter measurements \cite{Farooq}. Together, these include measurements up to redshift $z\sim2.36$. The Hubble constant is analytically marginalized using the prescription of \cite{Anagnostopoulos}.

Our astrophysical data consists of high-resolution spectroscopy tests of the stability of $\alpha$. We use a total of 319 measurements, of which 293 come from the analysis of archival data \cite{Webb} and the remaining 26 are dedicated measurements \cite{ROPP,Cooksey,Welsh,Milakovic}. These include measurements up to redshift $z\sim4.18$; the latter subset contains more stringent measurements, so overall the archival and dedicated subsets have comparable constraining power \cite{Meritxell}.

Additionally we use the geophysical constraint from the Oklo natural nuclear reactor \cite{Oklo}
\begin{equation} \label{okloalpha}
\frac{\Delta\alpha}{\alpha} =(0.5\pm6.1)\times10^{-8}\,,
\end{equation}
at an effective redshift $z=0.14$, though we note that underlying this bound is the simplifying assumption that $\alpha$ is the only parameter that may have been different and all the remaining physics is unchanged. As we have already mentioned, the current drift rate  of $\alpha$ is constrained by local comparison experiments between atomic clocks, with the most stringent bound being \cite{Lange}
\be
\frac{1}{H_0}\left(\frac{\dot\alpha}{\alpha}\right)_0=(1.4\pm1.5)\times10^{-8}\,.
\ee
Last but not least, we use the recent MICROSCOPE bound on the Eotvos parameter \cite{Touboul}
\be
\eta=(-0.1\pm1.3)\times10^{-14}\,,
\ee
which as previously discussed constrains the model's coupling to the electromagnetic sector.

We now use this data to constrain the model, using standard statistical analysis techniques. We consider two different parametrizations introduced in the previous section: the first focuses on the deviations of the equation of state from its cosmological constant value (and is therefore observationally motivated), while the second retains the theoretical model parameters, including the slope of the potential. The comparison of the two serves to probe the sensitivity of the obtained constraints to our choices of priors. Before this comparison, we start with an analysis including only the cosmological data, which serves as a benchmark for the constraining power of the astrophysical data.

\subsection{\label{part1}Constraints from cosmological data}

\begin{figure*}
\centering
    \includegraphics[width=1.0\columnwidth]{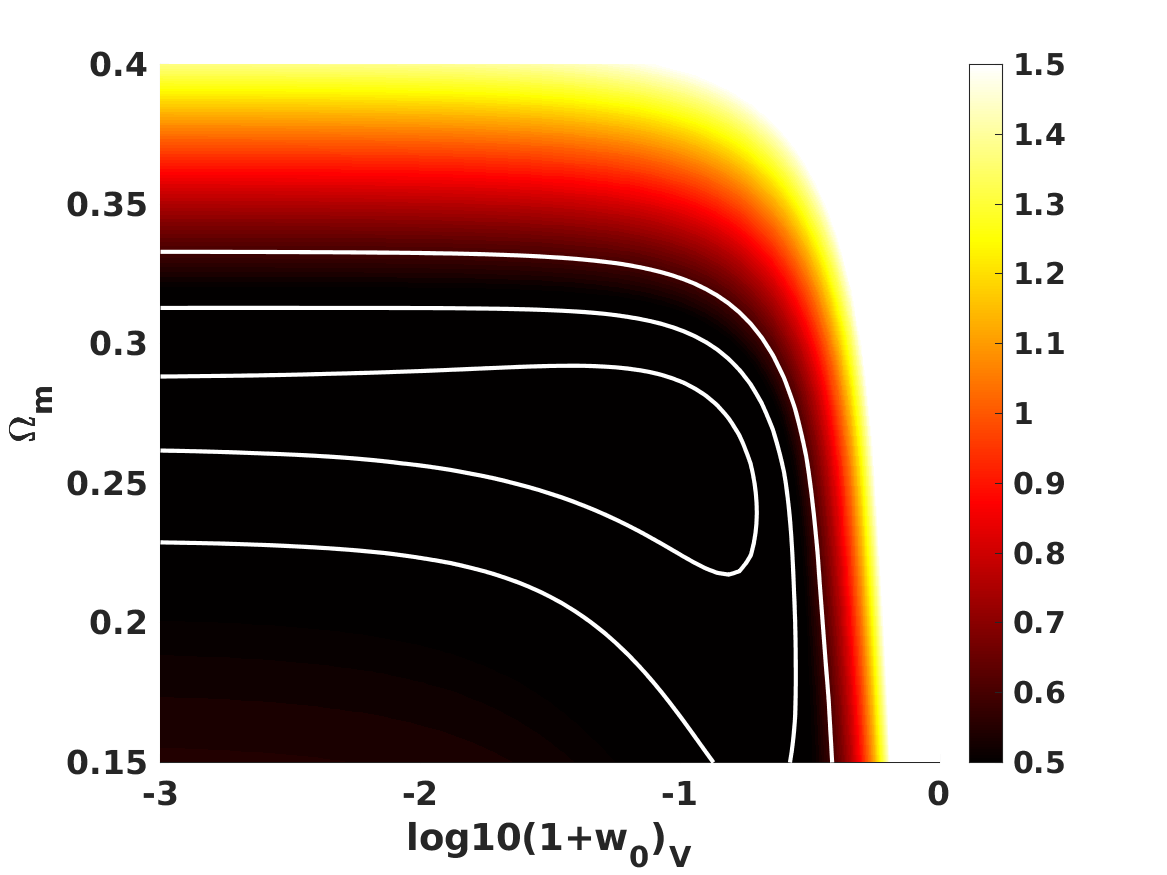}
    \includegraphics[width=1.0\columnwidth]{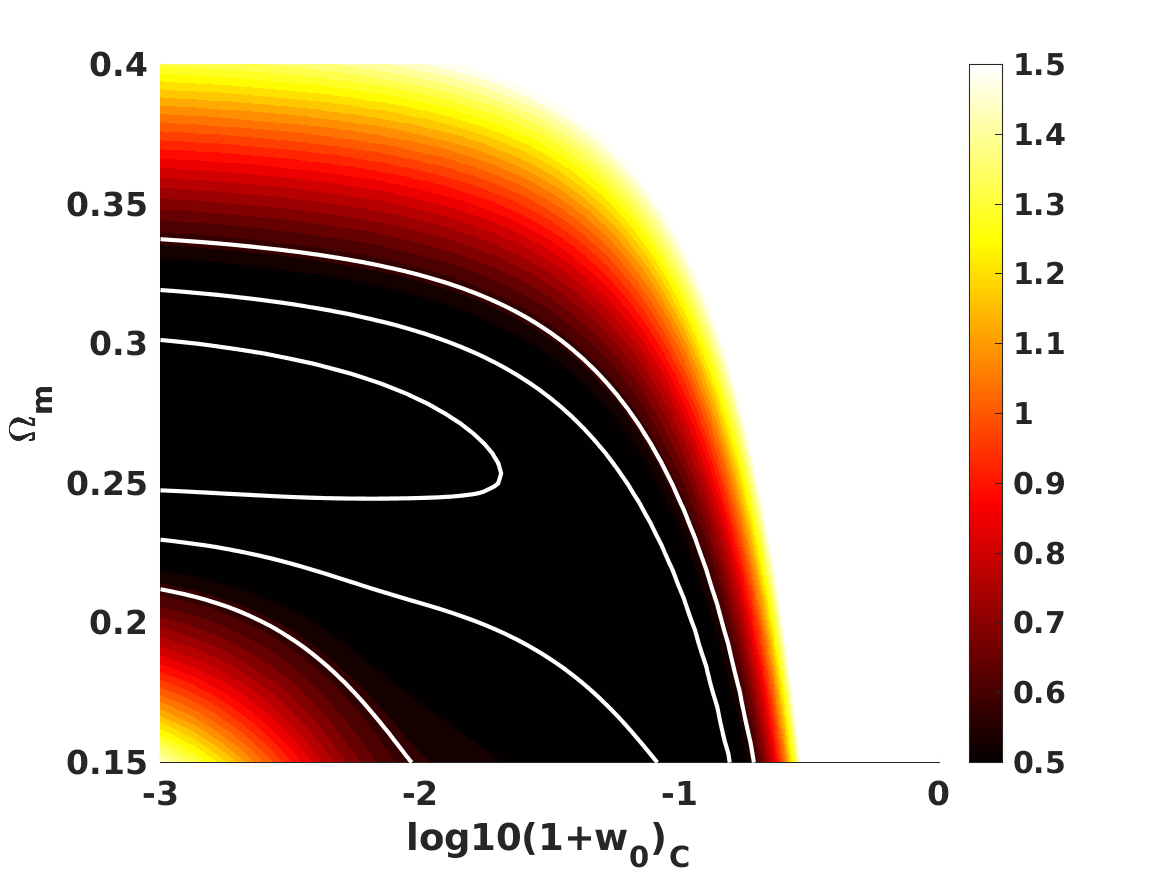}
    \includegraphics[width=1.0\columnwidth]{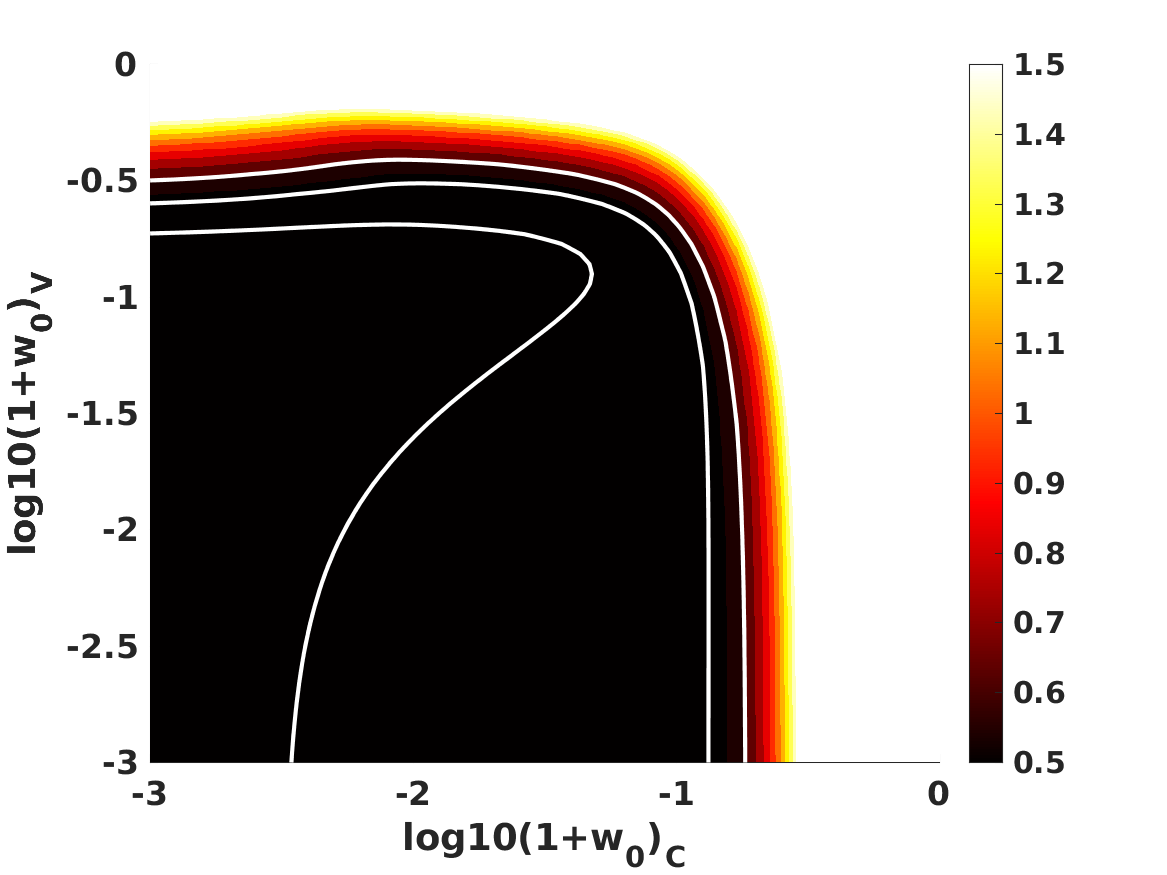}
    \includegraphics[width=1.0\columnwidth]{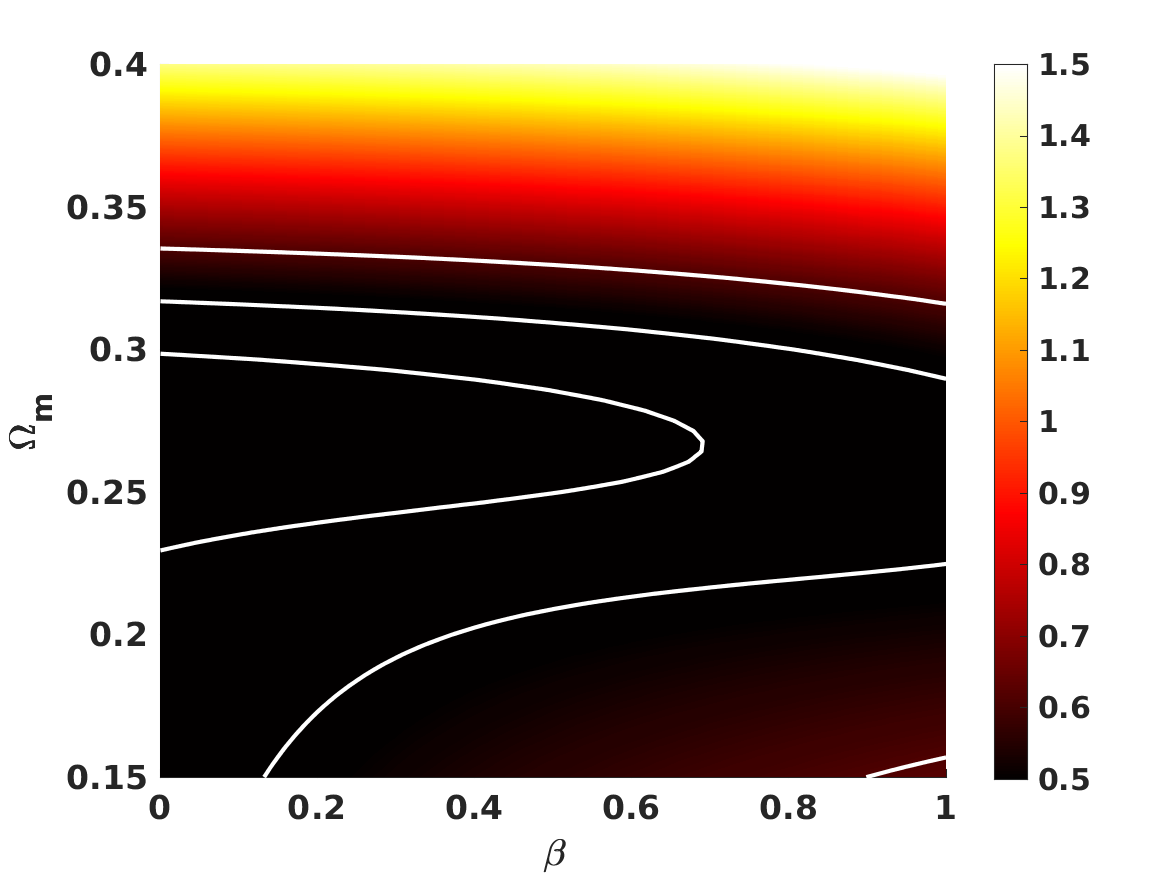}
    \includegraphics[width=1.0\columnwidth]{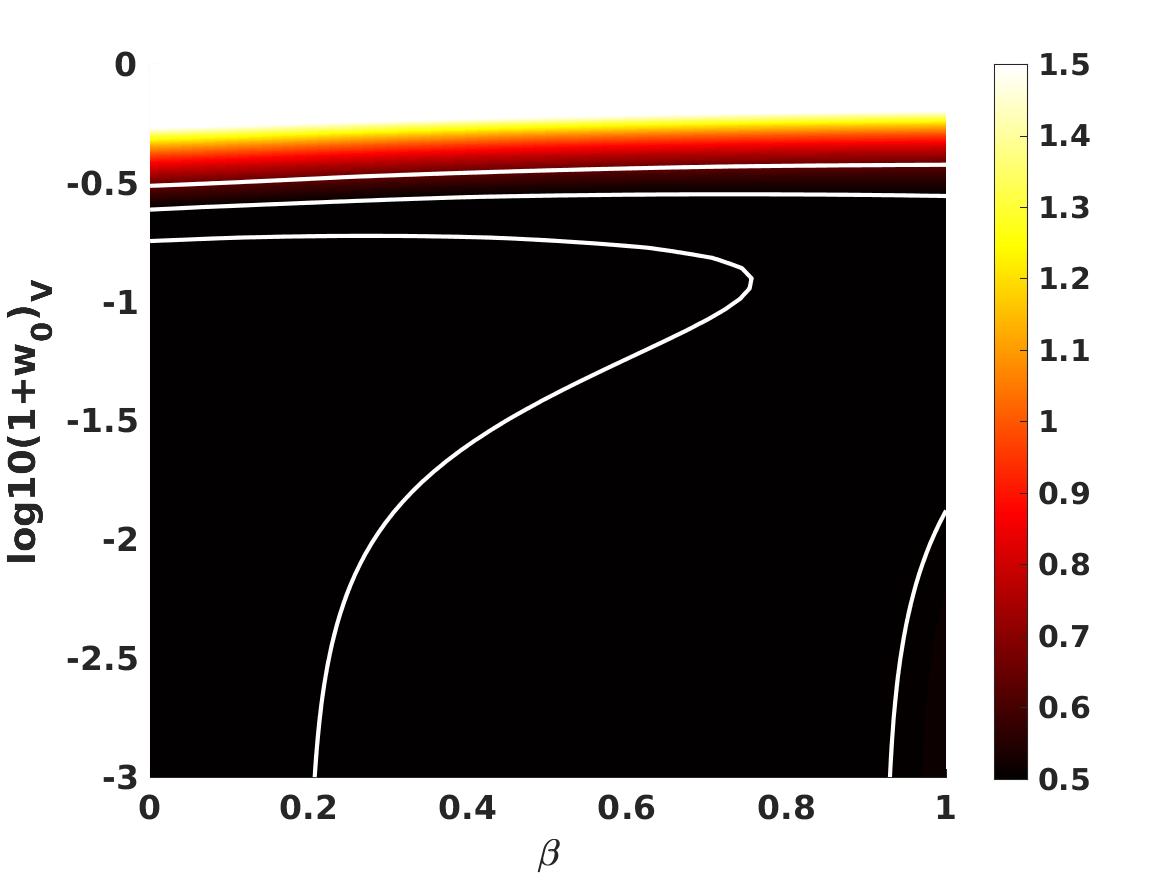}
    \includegraphics[width=1.0\columnwidth]{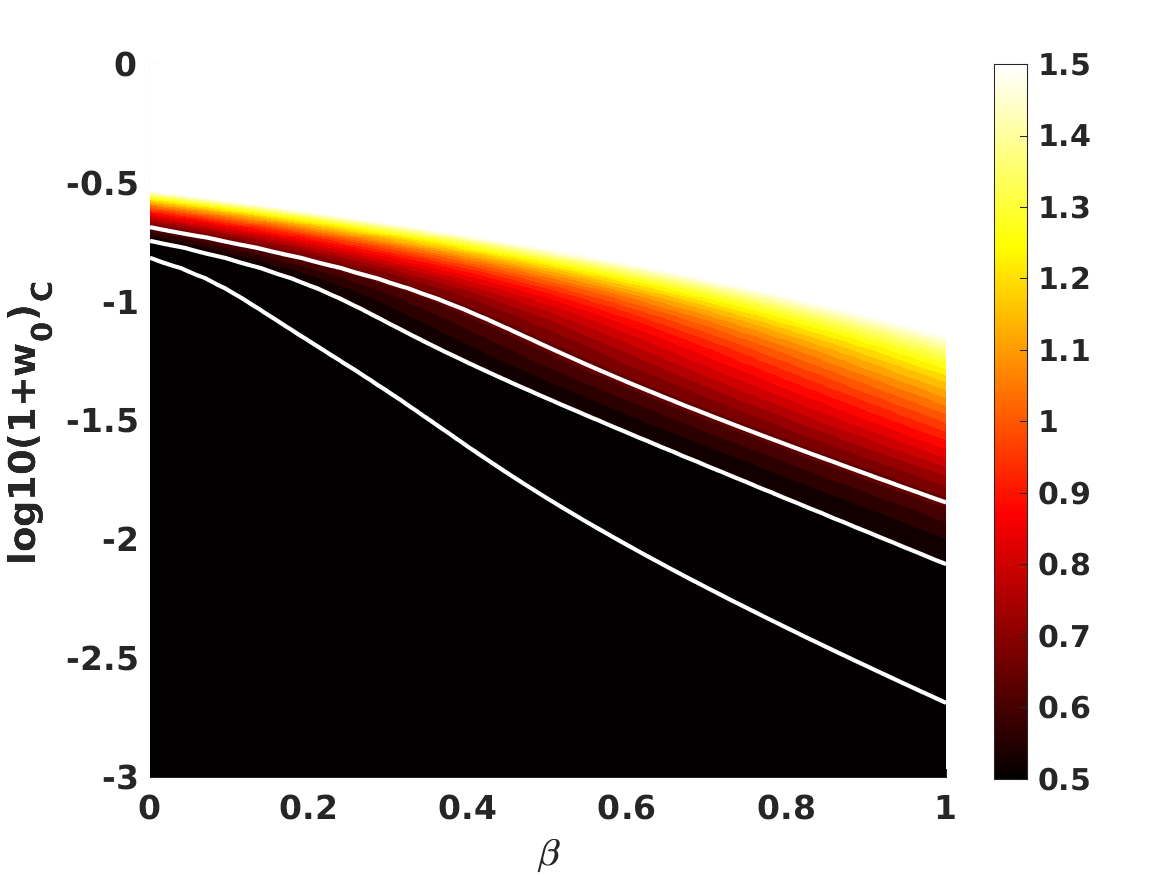}
  \caption{Constraints on generalized DBI models from our cosmological data, using the equation of state focused parametrization. The black contours denote the one, two, and three sigma confidence levels, and the colour map depicts the reduced $\chi^2$ of the fit for each set of model parameters (the white colour corresponds to a reduced $\chi^2$ of 1.5 or higher, and the black one to a reduced $\chi^2$ of 0.5 or lower).\label{fig1}}
\end{figure*}

\begin{figure*}
\centering
    \includegraphics[width=1.0\columnwidth]{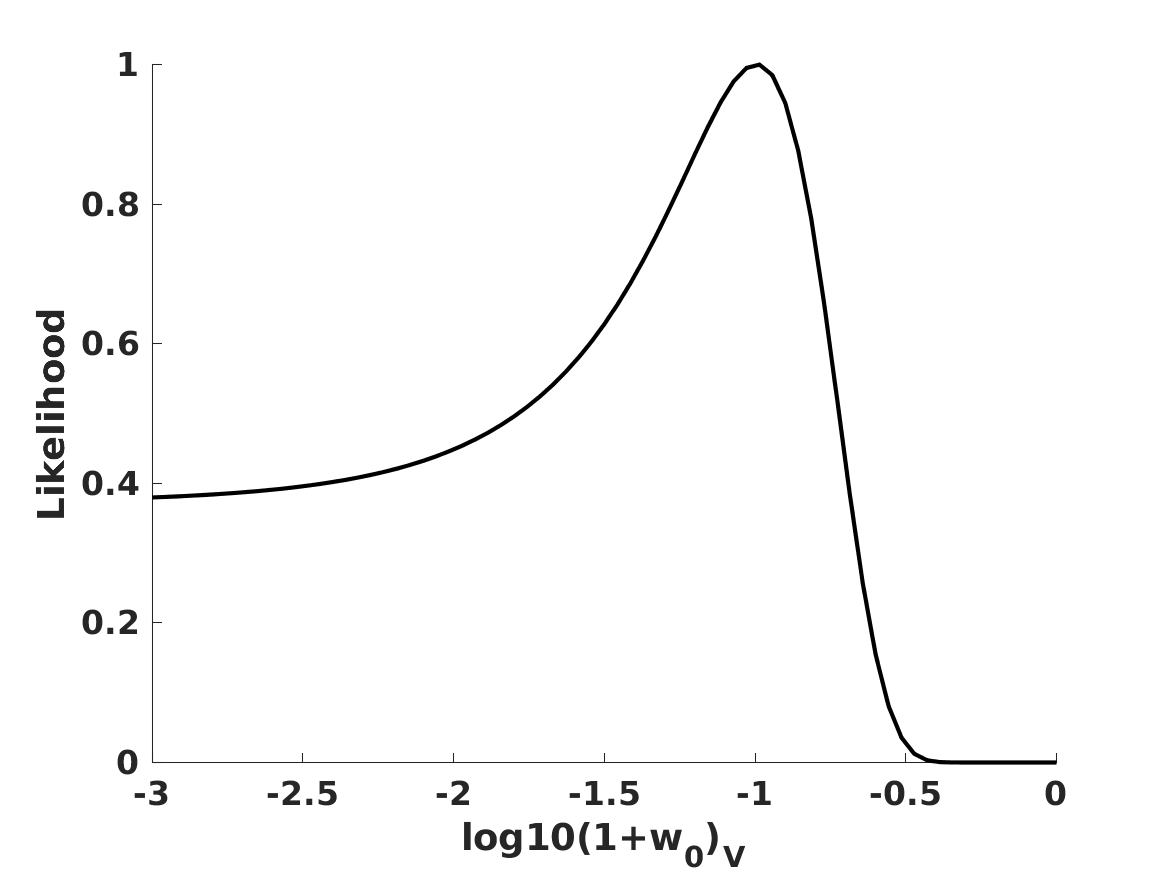}
    \includegraphics[width=1.0\columnwidth]{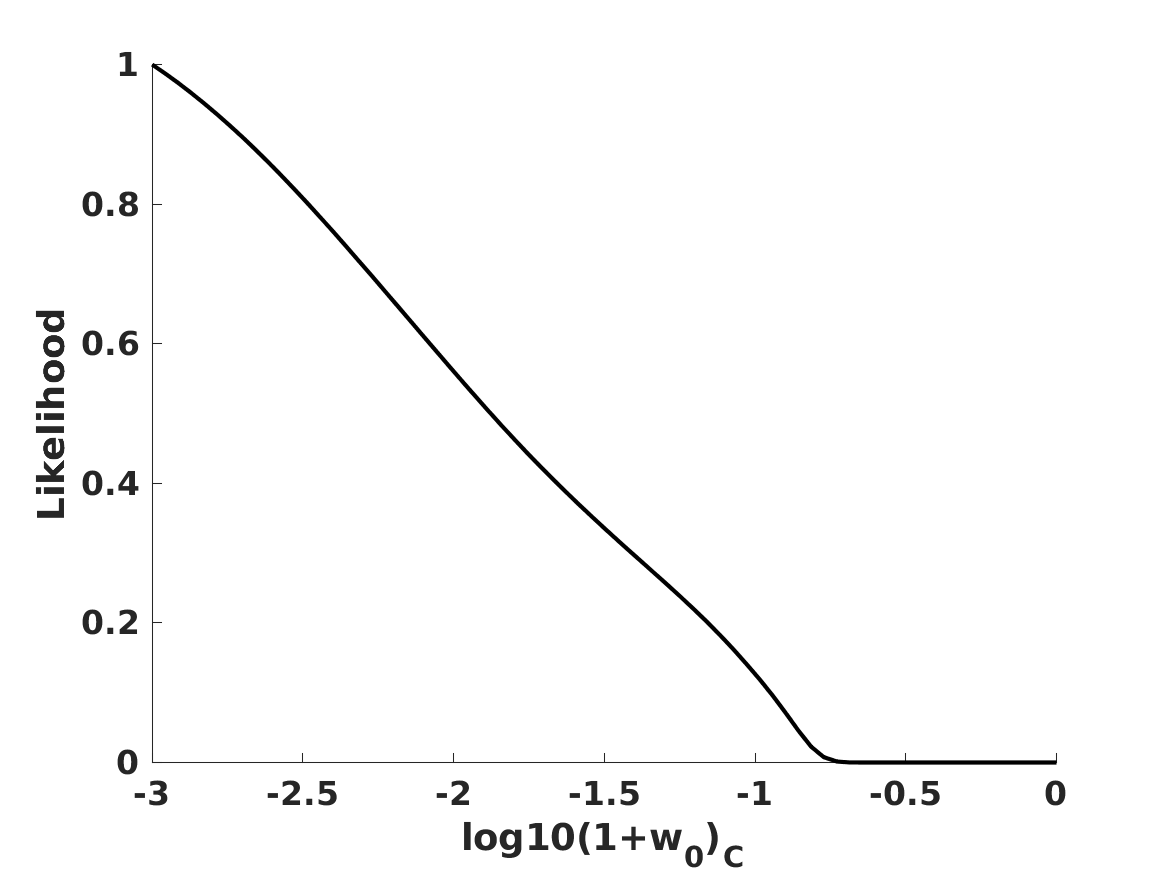}
    \includegraphics[width=1.0\columnwidth]{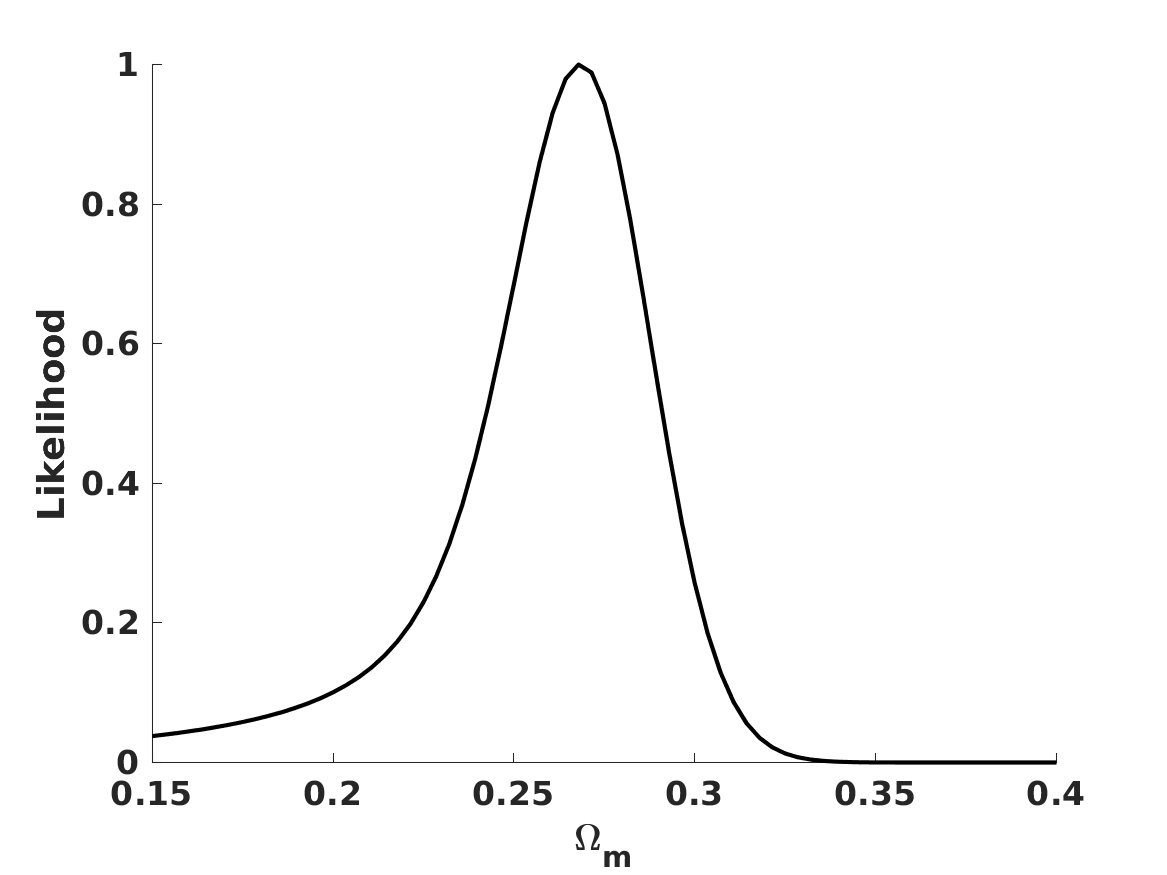}
    \includegraphics[width=1.0\columnwidth]{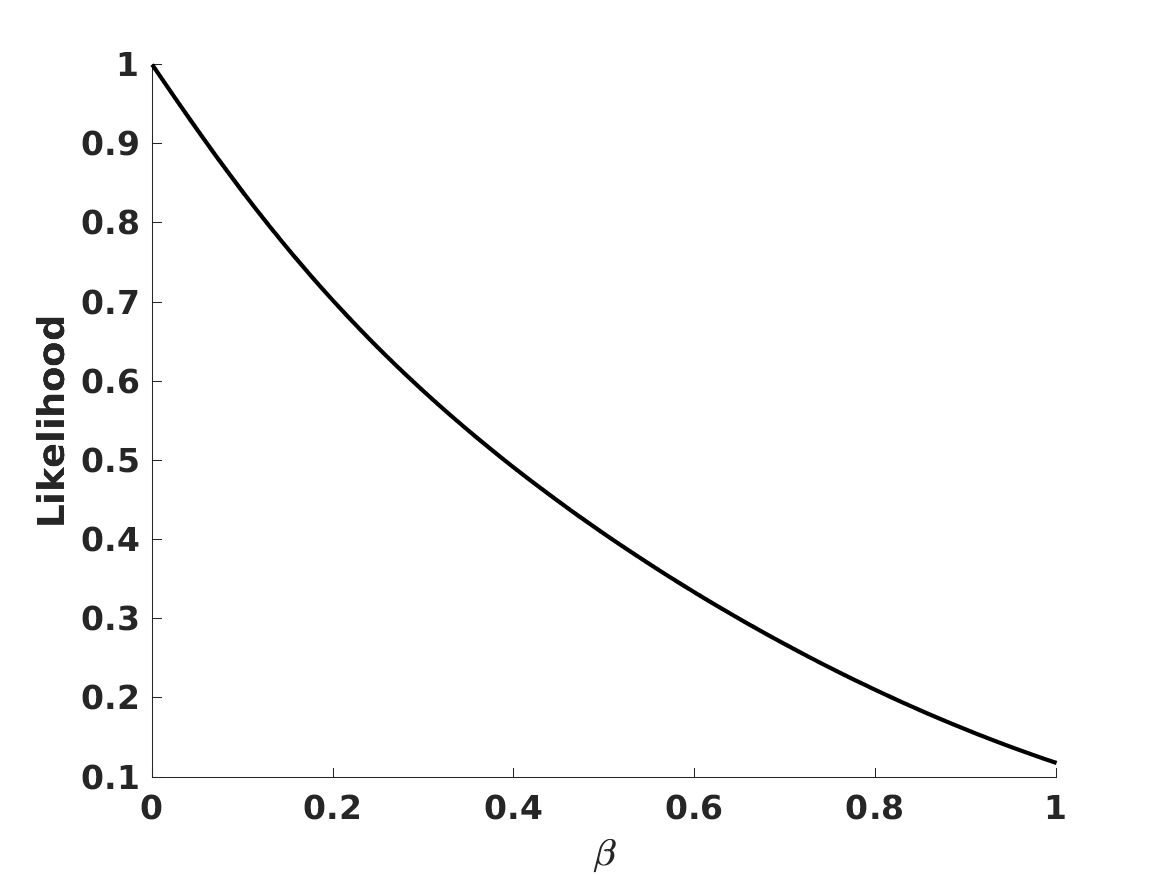}  
  \caption{One-dimensional (marginalized) posterior likelihoods for the model parameters corresponding to the analysis in Fig. \protect\ref{fig1}.\label{fig2}}
\end{figure*}

We assume that the four independent parameters are $(\log_{10}{(1+w_0)_V},\log_{10}{(1+w_0)_C},\Omega_m,\beta)$. We choose the following four uniform priors $\log_{10}{(1+w_0)_V}=[-3,0]$, $\log_{10}{(1+w_0)_C}=[-3,0]$, $\Omega_m=[0.15,0.40]$ and $\beta=[0,1]$. The results of the analysis including only the aforementioned supernova and Hubble parameter data are depicted in Figs. \ref{fig1} and \ref{fig2}. The reduced chi-square at the four-dimensional best fit model is $\chi^2_\nu=0.67$, which (unsurprisingly) indicates that the data is overfitting this four-parameter model. Nevertheless, for the purpose of comparison with the analysis of the full datasets, we note that the posterior likelihood for the matter density is
\be
\Omega_m=0.27^{+0.04}_{-0.06}\,
\ee
while for the equation of state parameters we obtain the upper bounds
\be
\log_{10}{(1+w_0)_V}<-0.59
\ee
\be
\log_{10}{(1+w_0)_C}<-1.01\;
\ee
all of the above are given at the two sigma ($95.4\%$) confidence level.

For the parameter $\beta$ we can only get the weak one sigma ($68.3\%$) confidence level bound
\be
\beta<0.28\,,
\ee
but the parameter is unconstrained at the two sigma level. A low value of $\beta$ is preferred since it leads to a weaker redshift dependence of the dark energy component in the Friedmann equation. In this case, the model's dark energy equation of state $(1+w_{eff})$ is therefore allowed significant deviations from the cosmological constant behaviour which may come from both the tachyon and the Chaplygin mechanisms.

\subsection{\label{part2}Constraining the dark energy equation of state}

Here we assume that the four independent parameters are the same as in the previous subsection, but change the priors in the two dark energy parameters to $\log_{10}{(1+w_0)_V}=[-10,-4]$ and  $\log_{10}{(1+w_0)_C}=[-4,0]$, while those on $\Omega_m$ and $\beta$ remain unchanged. Note that since we are including the astrophysical (and local) data in the analysis it is obvious a priori that the $\alpha$ constraints will require the value of $\log_{10}{(1+w_0)_V}$ to be significantly negative, while there is no equally stringent constraint for $\log_{10}{(1+w_0)_C}$.

\begin{figure*}
\centering
    \includegraphics[width=1.0\columnwidth]{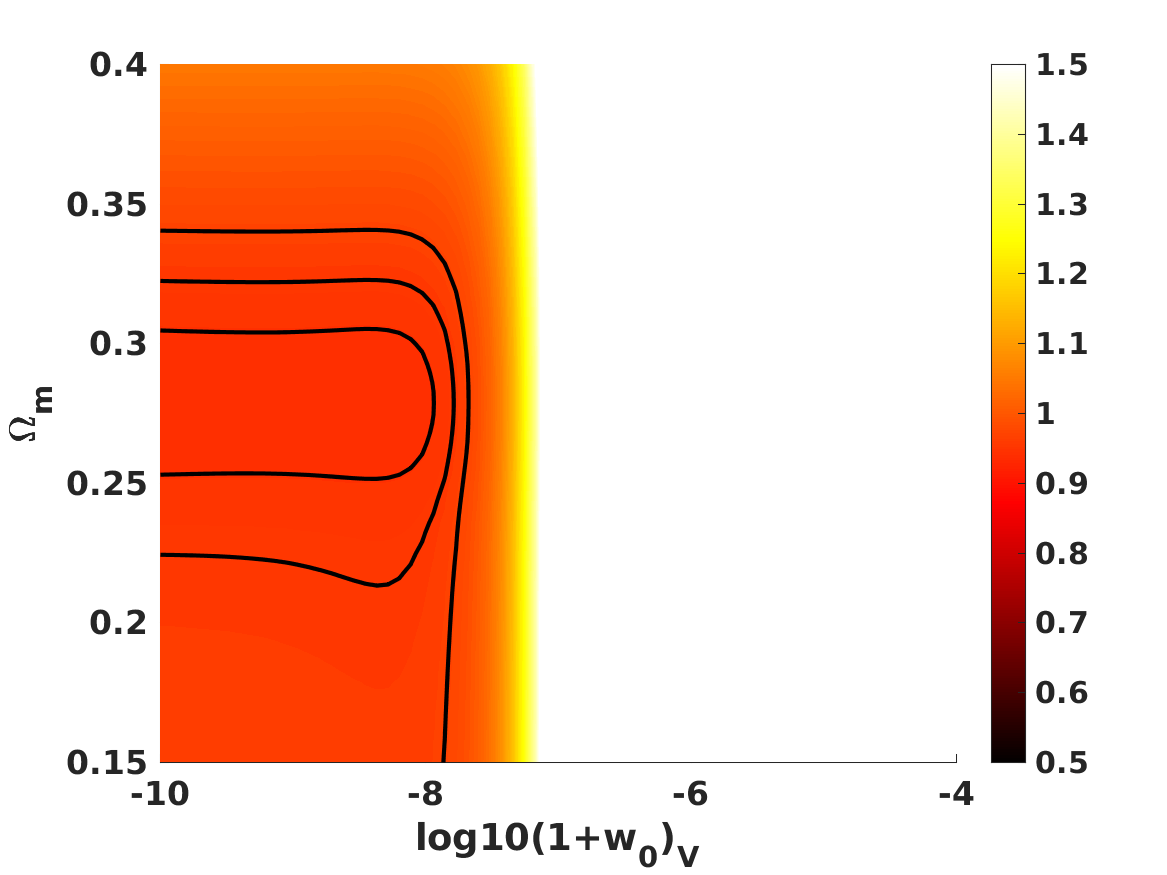}
    \includegraphics[width=1.0\columnwidth]{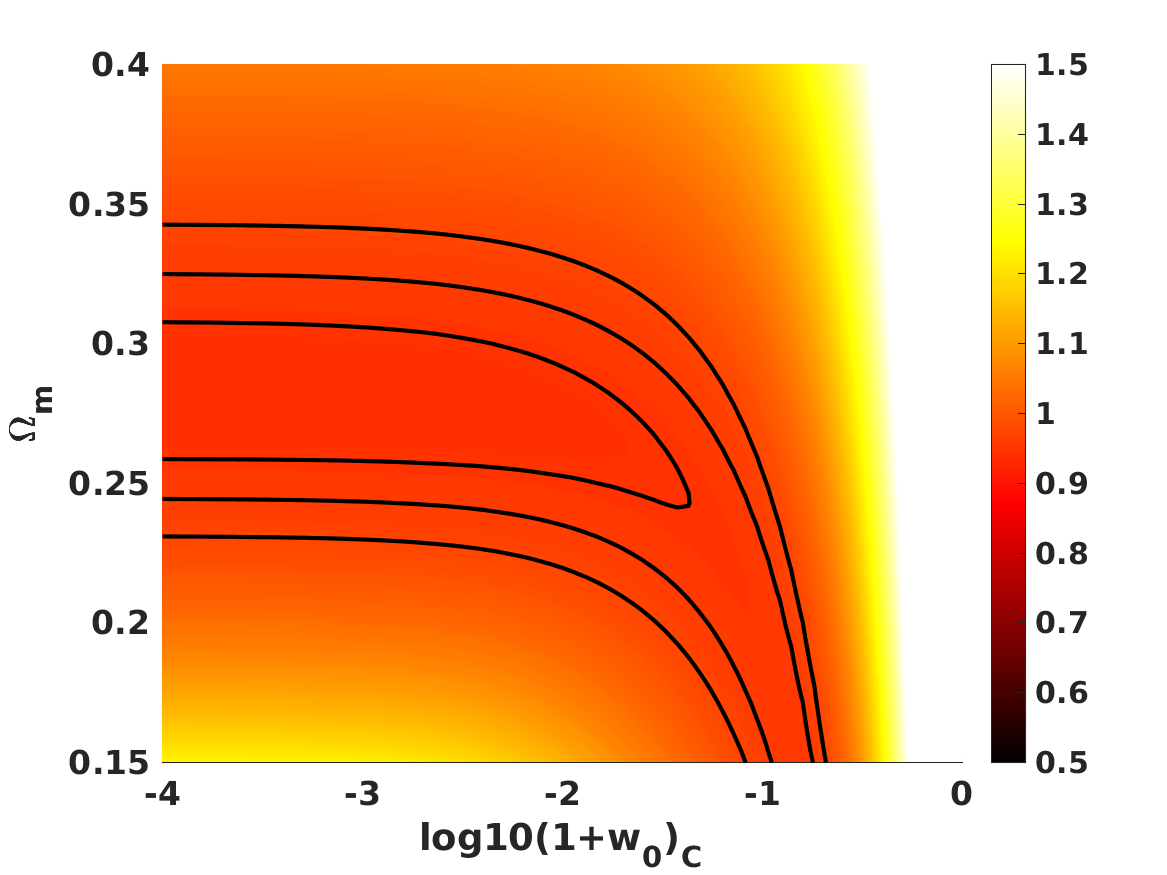}
    \includegraphics[width=1.0\columnwidth]{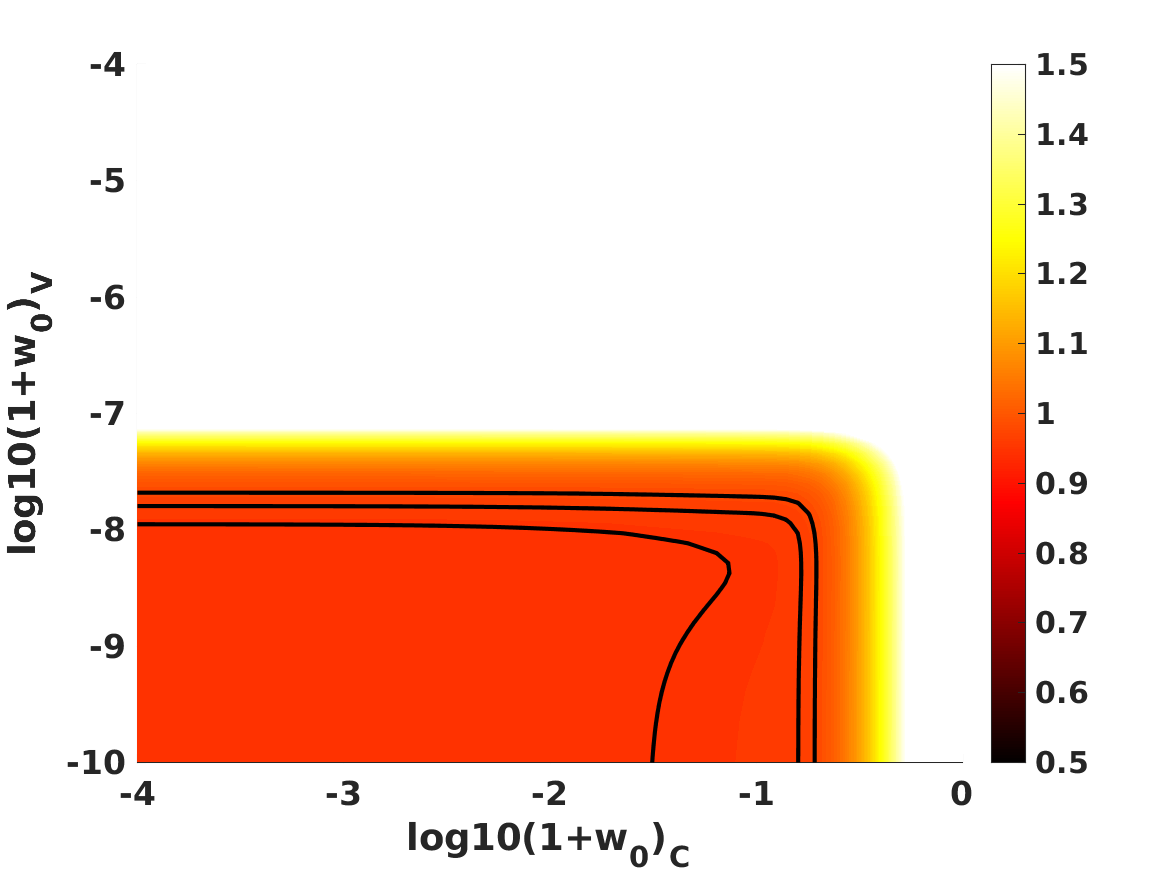}
    \includegraphics[width=1.0\columnwidth]{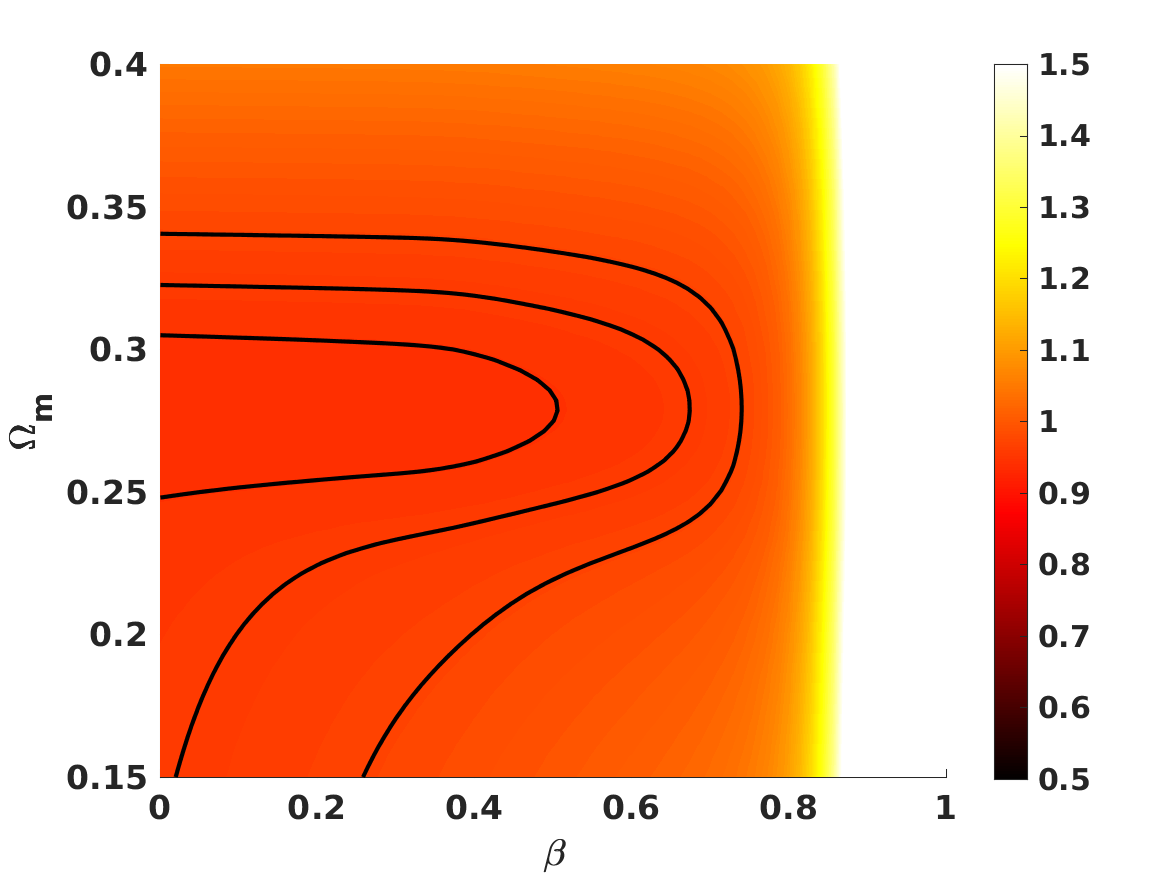}
    \includegraphics[width=1.0\columnwidth]{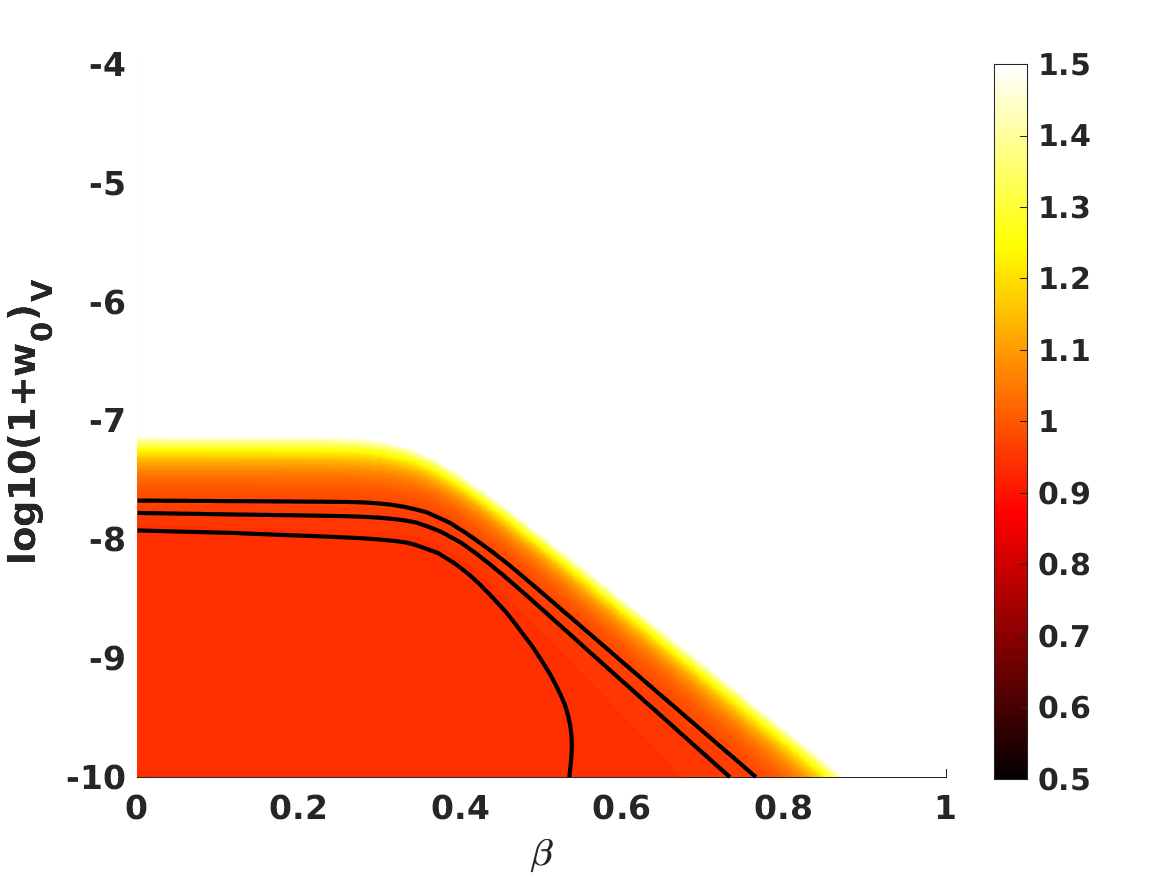}
    \includegraphics[width=1.0\columnwidth]{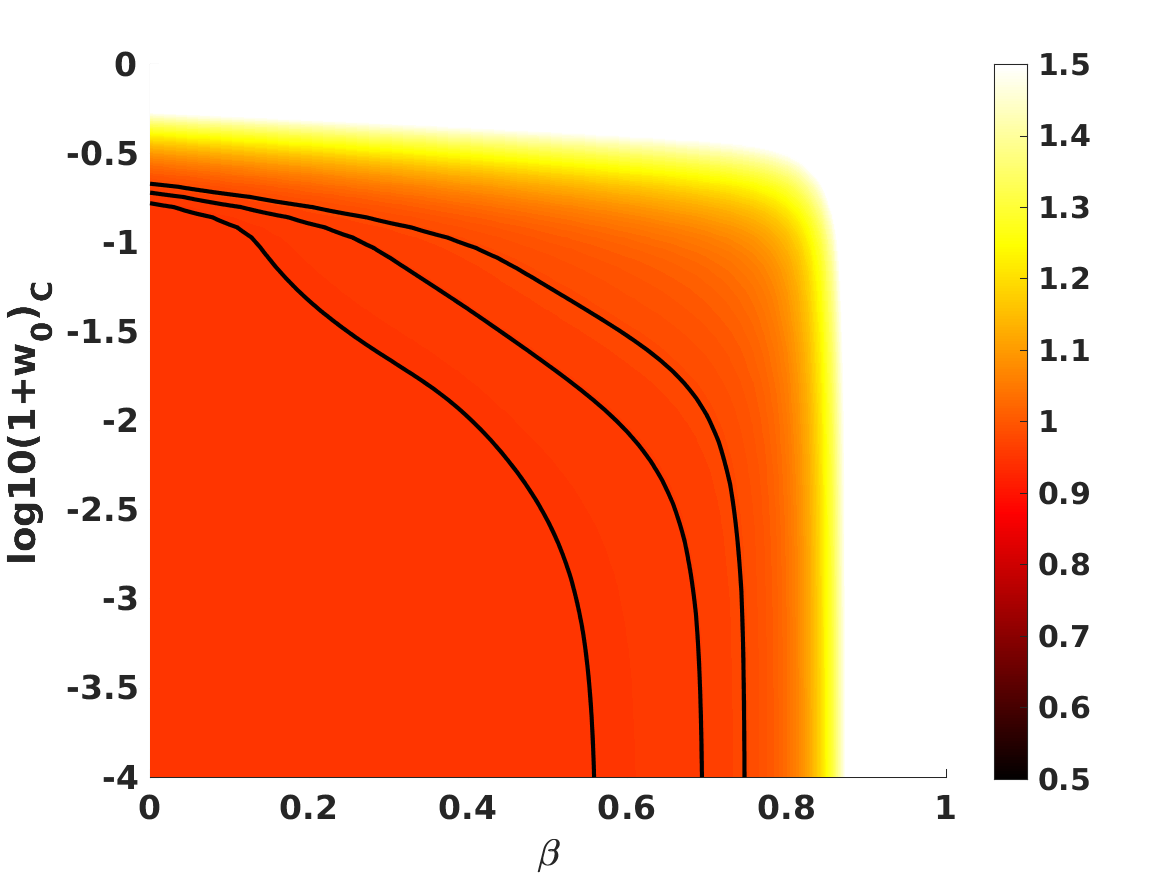}
  \caption{Constraints on generalized DBI models, using the equation of state focused parametrization. The black contours denote the one, two, and three sigma confidence levels, and the colour map depicts the reduced $\chi^2$ of the fit for each set of model parameters (the white colour corresponds to a reduced $\chi^2$ of 1.5 or higher).\label{fig3}}
\end{figure*}

\begin{figure*}
\centering
    \includegraphics[width=1.0\columnwidth]{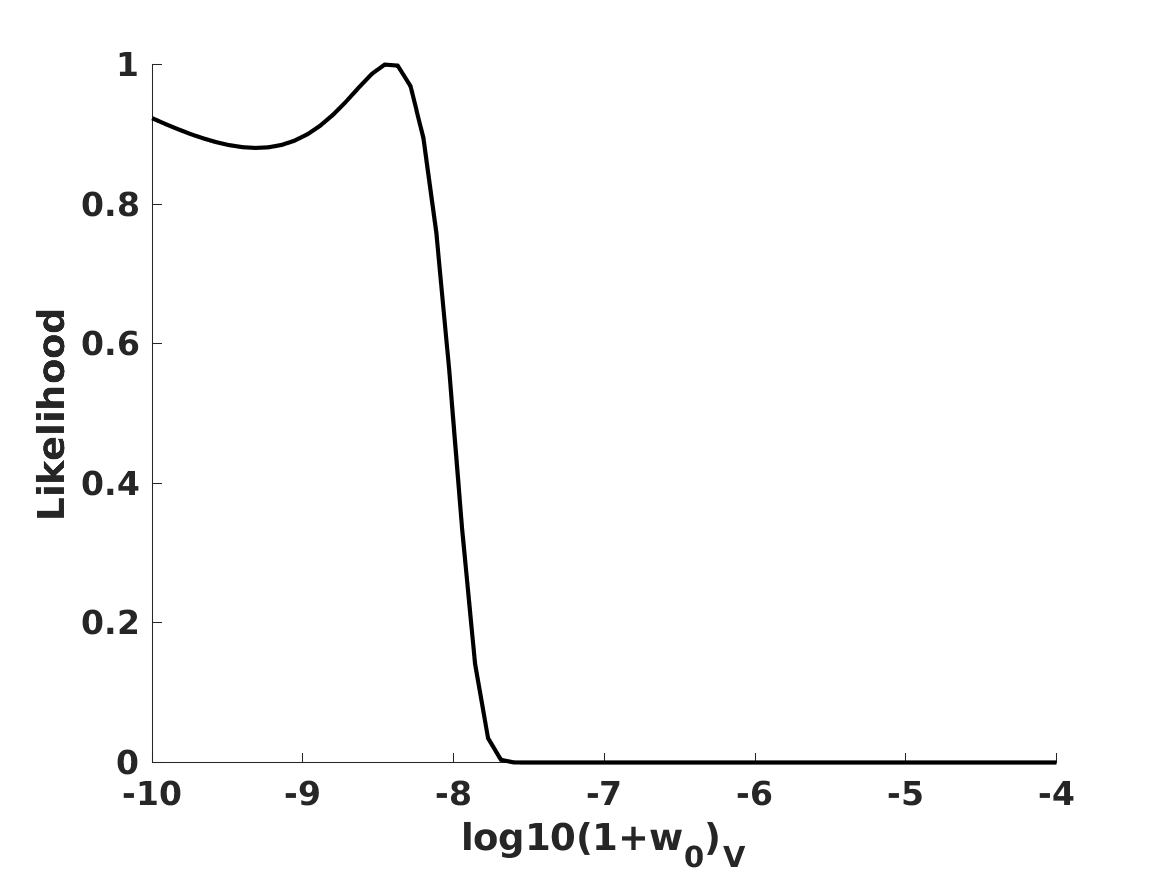}
    \includegraphics[width=1.0\columnwidth]{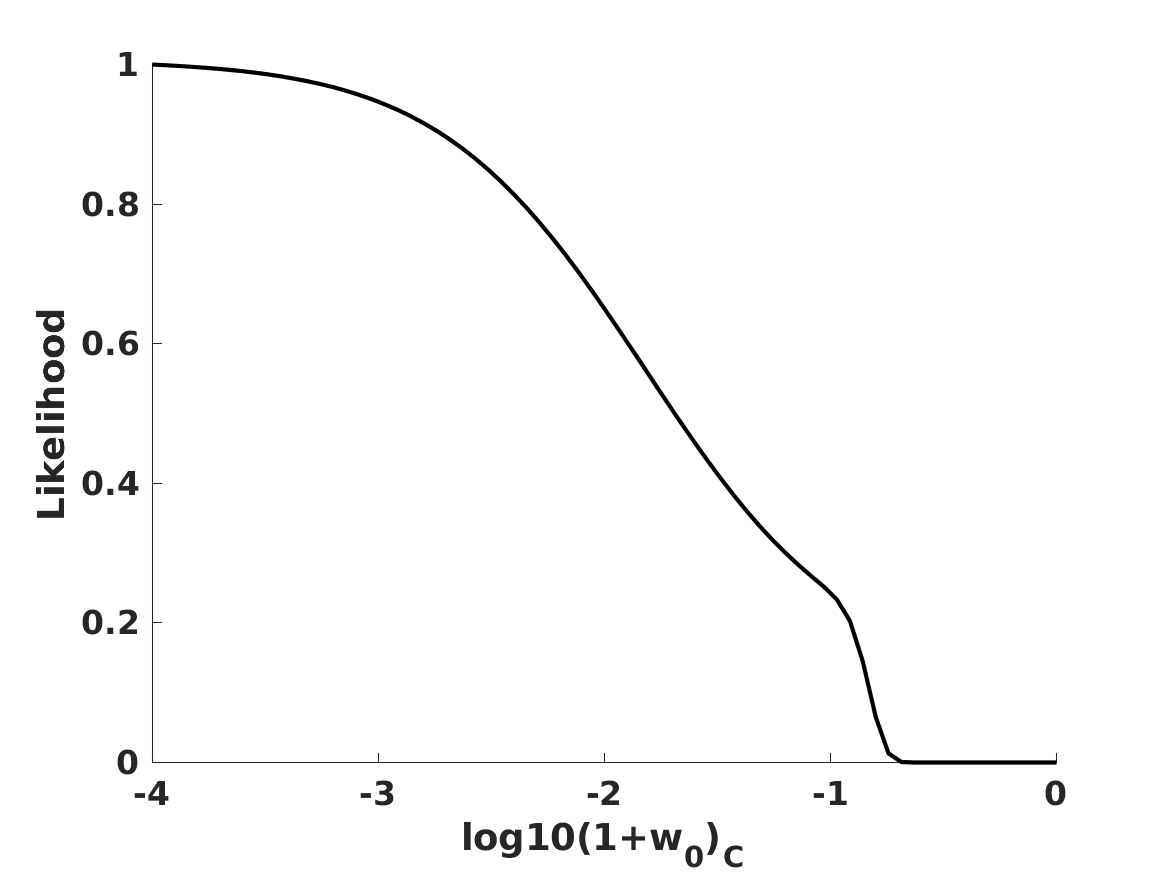}
    \includegraphics[width=1.0\columnwidth]{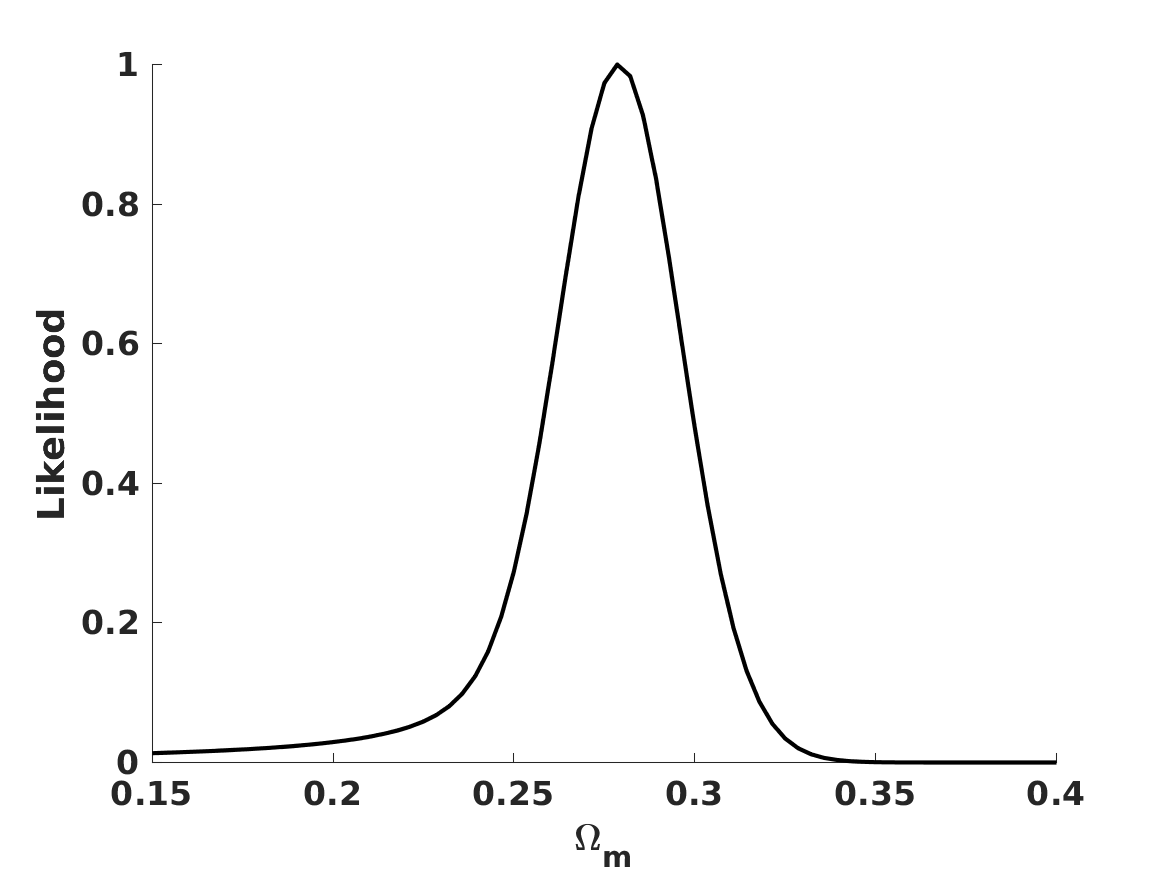}
    \includegraphics[width=1.0\columnwidth]{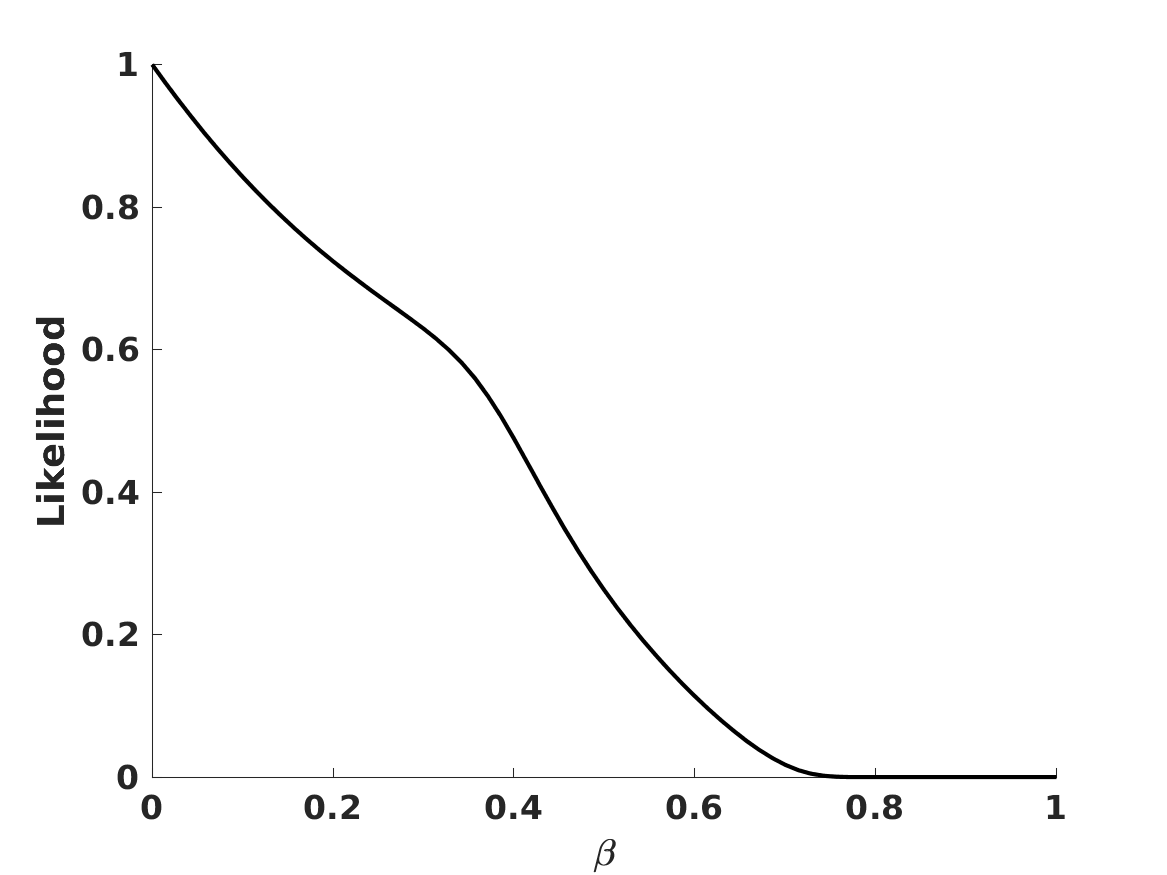}  
  \caption{One-dimensional (marginalized) posterior likelihoods for the model parameters corresponding to the analysis in Fig. \protect\ref{fig3}.\label{fig4}}
\end{figure*}

The results of this analysis are depicted in Figs. \ref{fig3} and \ref{fig4}. The reduced chi-square at the four-dimensional best fit model is now $\chi^2_\nu=0.98$, corresponding to a very reasonable fit. As for the posterior likelihoods for the individual model parameters, the matter density is now constrained to be
\be
\Omega_m=0.28\pm0.04\,,
\ee
while for the equation of state parameters we obtain
\be
\log_{10}{(1+w_0)_V}<-7.85
\ee
\be
\log_{10}{(1+w_0)_C}<-0.85\;
\ee
all of these are still at the two sigma ($95.4\%$) confidence level. As expected, the rolling tachyon (i.e., potential slope) contribution is constrained to be extremely close to a cosmological constant (in other words, a flat potential) due to the astrophysical and atomic clock measurements, while the cosmological data constrains the remaining parameters. The constraint to the Chaplygin part is slightly weakened, while the preferred matter density constraint is slightly changed. The first of these shifts is due to the correlation of this parameter with $\beta$ (whose likelihood changes, as discussed in the next paragraph) while the second is due to the breaking of the degeneracy for low values of this density which exists when only the cosmological data is used.

The parameter $\beta$ is now more constrained, particularly at high values, and now we do get a two sigma bound
\be
\beta<0.58\,.
\ee
Again the low $\beta$ is preferred due to its weaker redshift dependence of the dark energy component in the Friedmann equation---recall that in this parametrization the fine-structure constant $\alpha$ does not depend on $\beta$. Moreover, larger values of $\beta$ are more sensitive to the Eotvos parameter constraint: in the $\beta=0$ limit one has $\eta\propto (1+w_0)_V^2$, while in the $\beta=1$ limit one has $\eta\propto (1+w_0)_V$ Thus the main outcome of the analysis for this parametrization is that the model's dark energy equation of state is effectively given by $(1+w_{eff})\simeq(1+w_0)_C$, since the astrophysical measurements tightly constrain the rolling tachyon part.

\subsection{\label{part3}Constraining the potential slope}

Here, we assume that the four independent parameters are $(\log_{10}{\lambda},\log_{10}{(1+w_0)_C},\Omega_m,\beta)$. For the potential slope we choose the uniform prior $\log_{10}{\lambda}=[-10,-4]$, while the priors for the other three parameters are the same as in the previous subsection. As before, it is a priori clear that the dimensionless slope of the potential must be significantly smaller than unity---if nothing else, from the previous results in \cite{Roll}.

Note that we expect significant differences between this parametrization and the previous one. The main reason is that the relation between $(1+w_0)_V$ and $\lambda$ also involves $\beta$, c.f. Eq. \ref{vbetalambda}. Thus when replacing $(1+w_0)_V$ by $\lambda$ in our analysis we are affecting the structure of the degeneracies between the remaining parameters in the corresponding parameter spaces, which will be reflected in the constraints on the common model parameters.

\begin{figure*}
\centering
    \includegraphics[width=1.0\columnwidth]{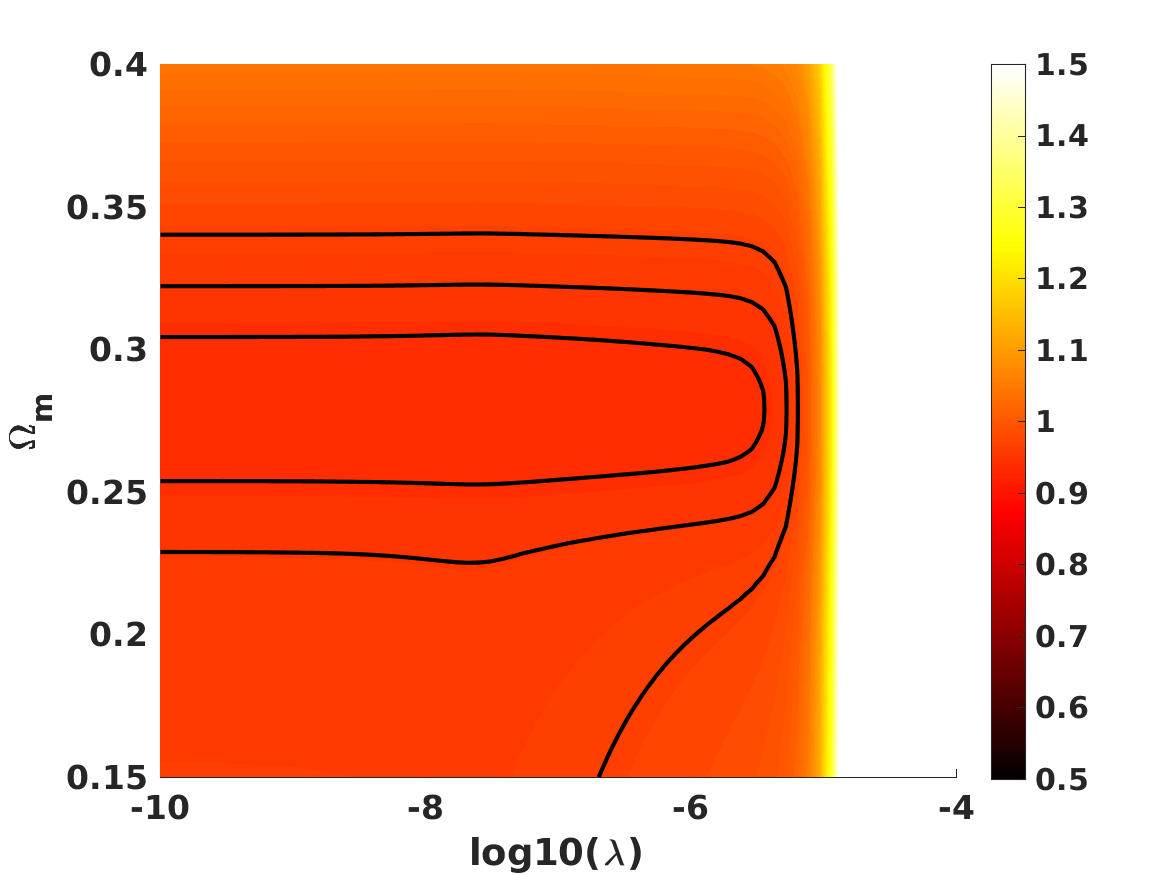}
    \includegraphics[width=1.0\columnwidth]{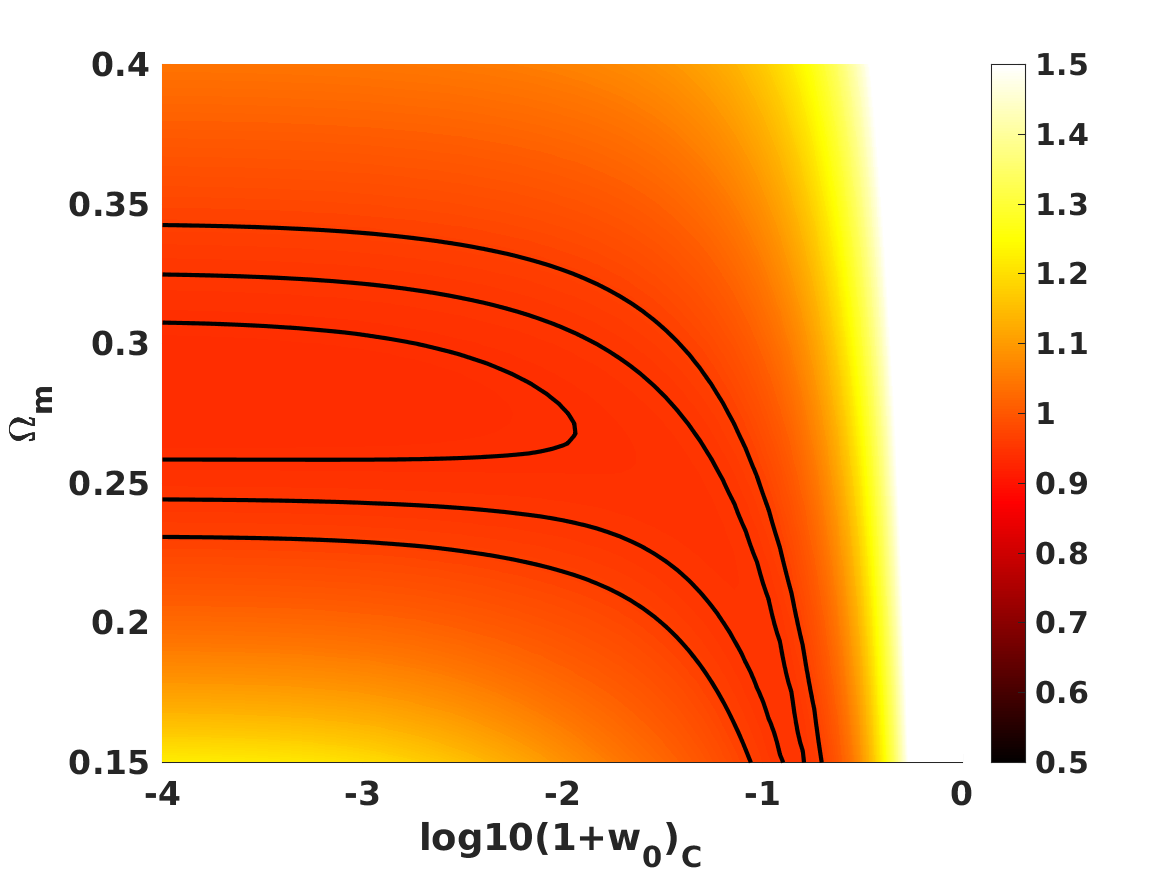}
    \includegraphics[width=1.0\columnwidth]{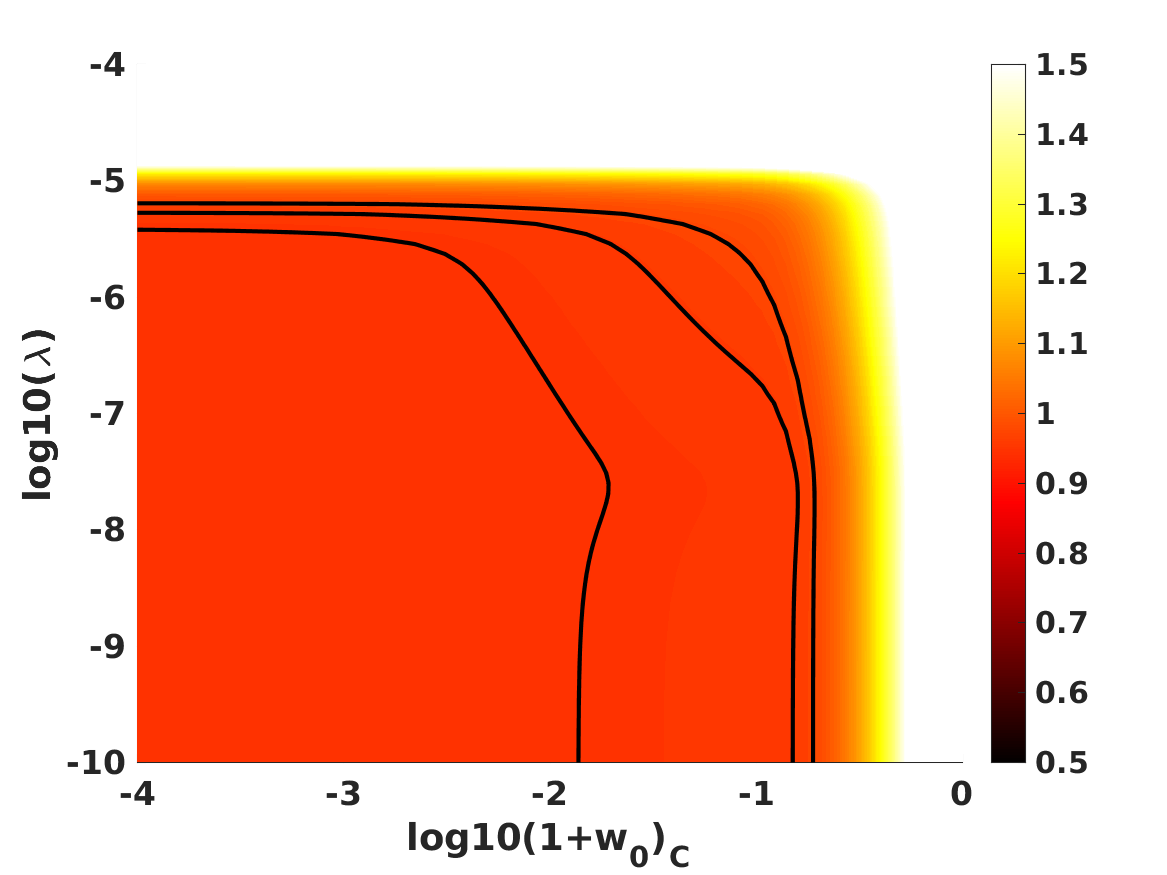}
    \includegraphics[width=1.0\columnwidth]{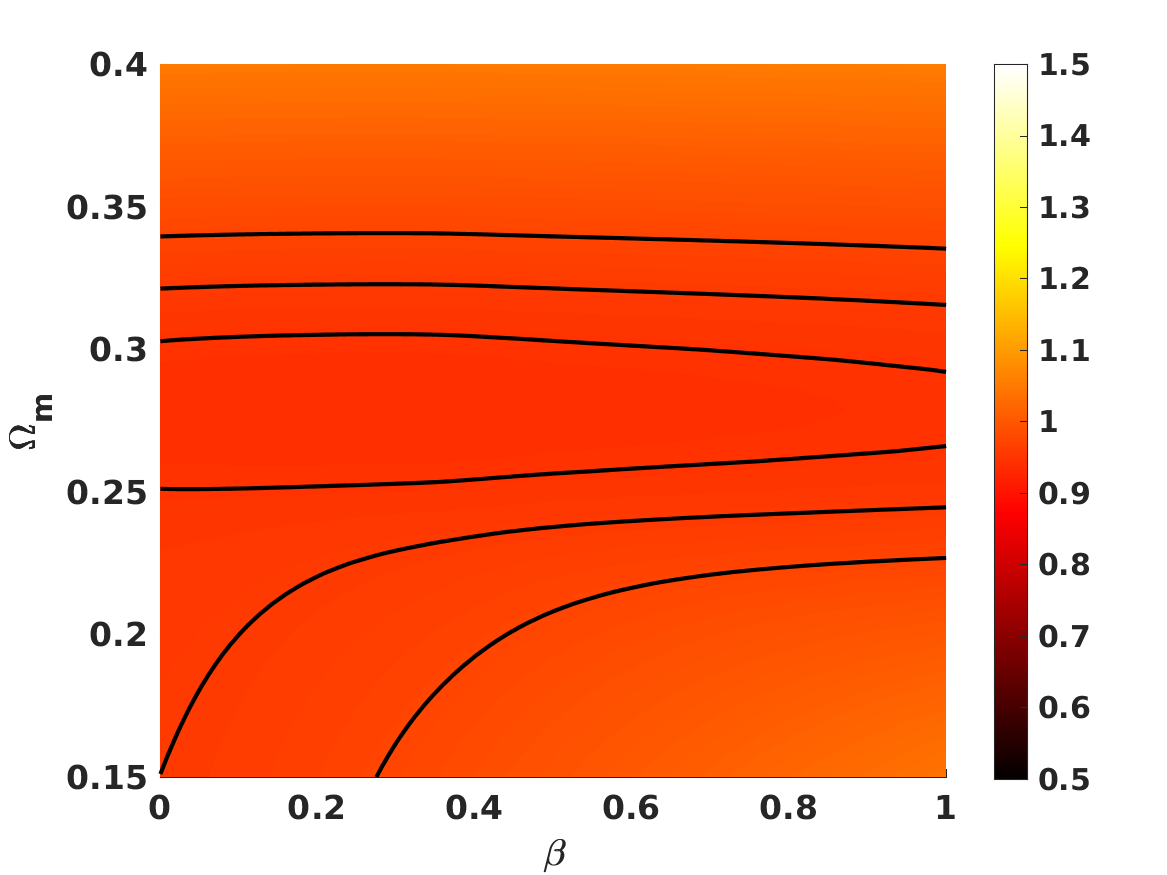}
    \includegraphics[width=1.0\columnwidth]{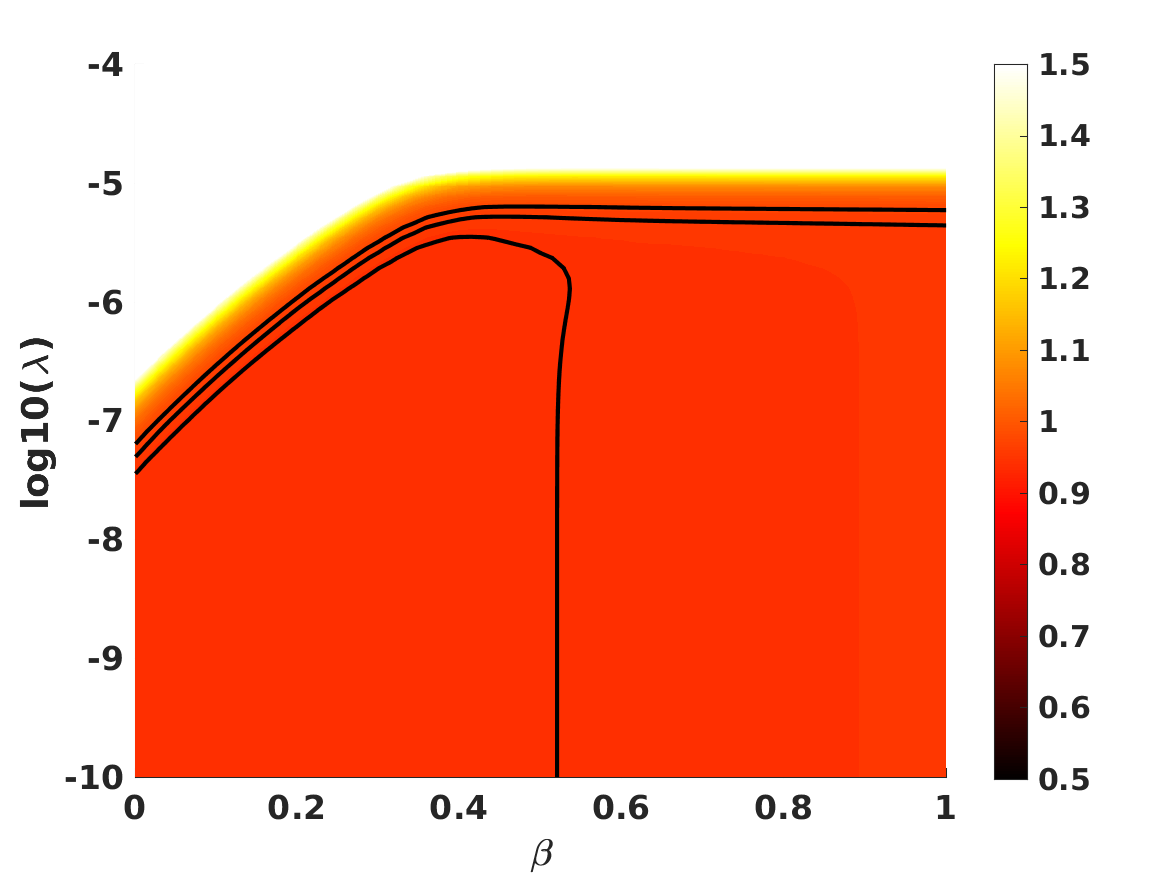}
    \includegraphics[width=1.0\columnwidth]{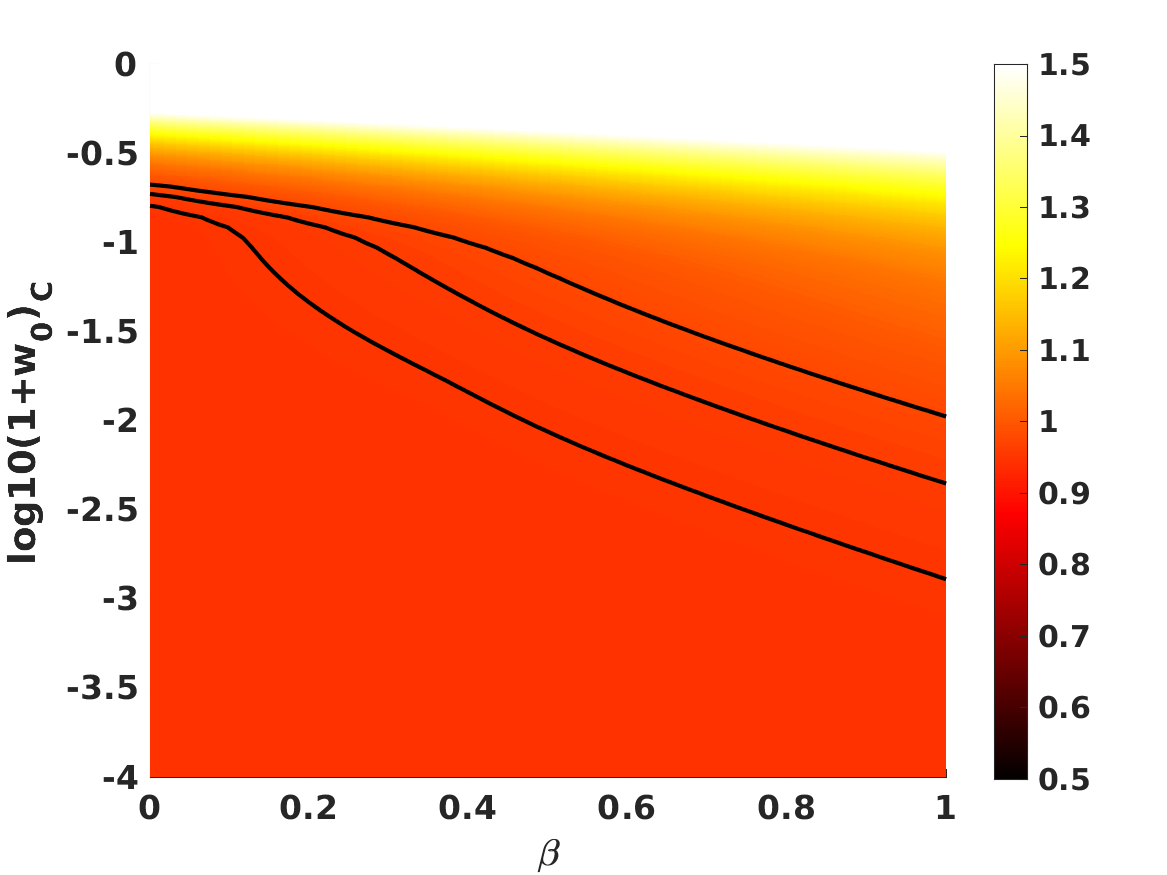}
  \caption{Constraints on generalized DBI models, using the potential slope parametrization. The black contours denote the one, two, and three sigma confidence levels, and the colour map depicts the reduced $\chi^2$ of the fit for each set of model parameters (the white colour corresponds to a reduced $\chi^2$ of 1.5 or higher).\label{fig5}}
\end{figure*}

\begin{figure*}
\centering
    \includegraphics[width=1.0\columnwidth]{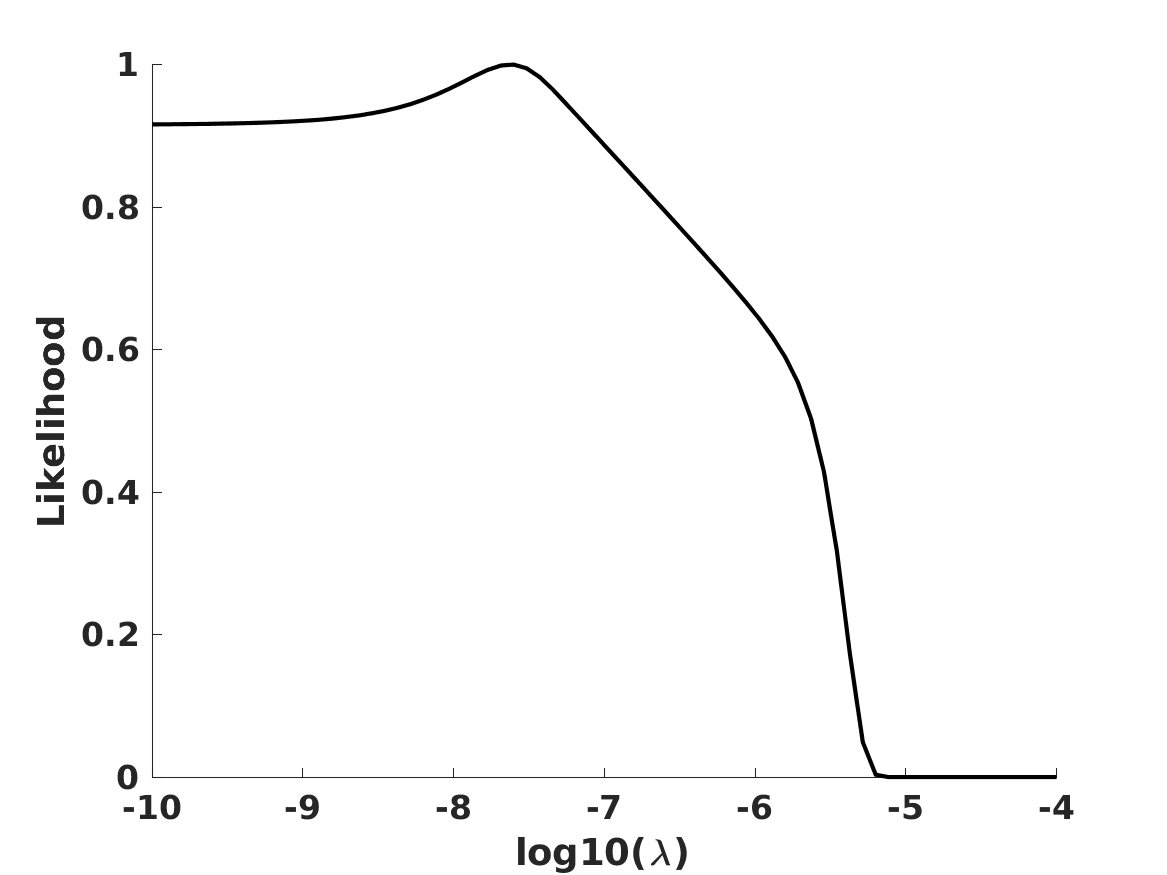}
    \includegraphics[width=1.0\columnwidth]{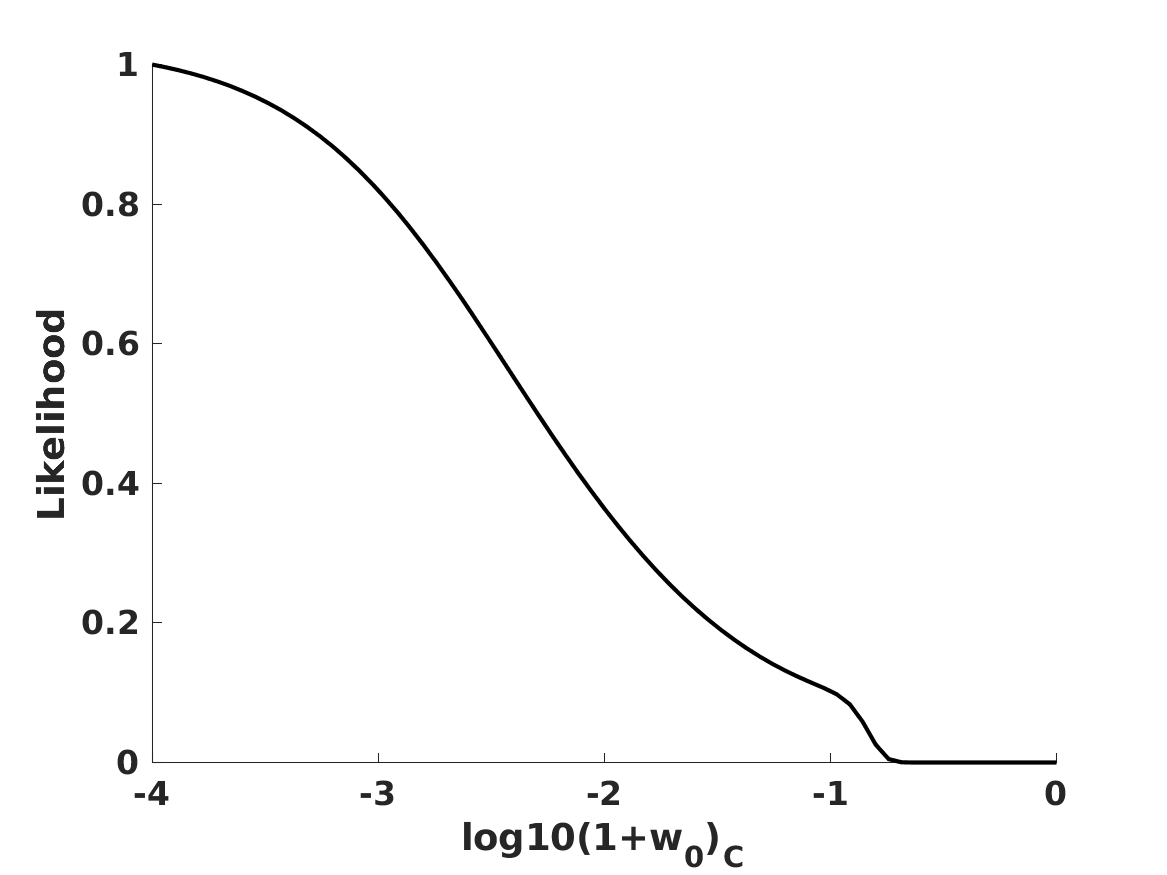}
    \includegraphics[width=1.0\columnwidth]{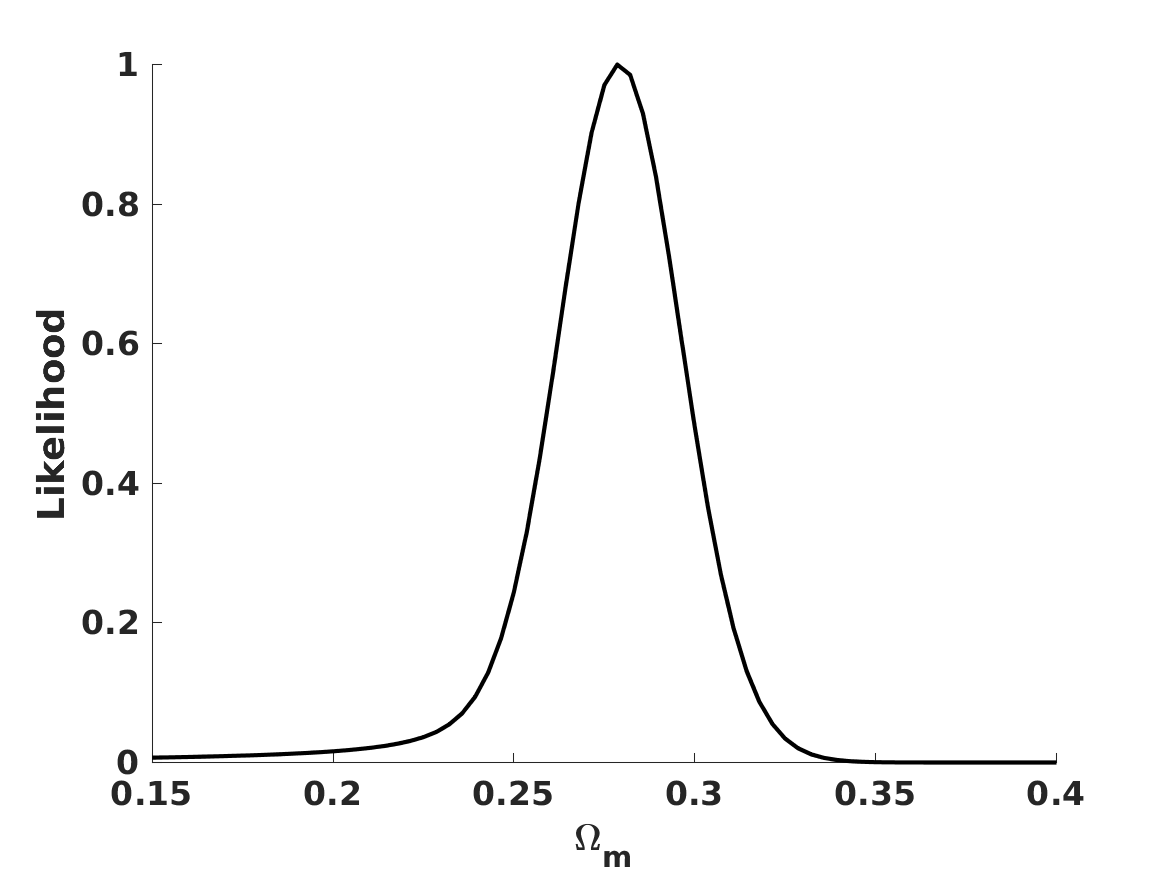}
    \includegraphics[width=1.0\columnwidth]{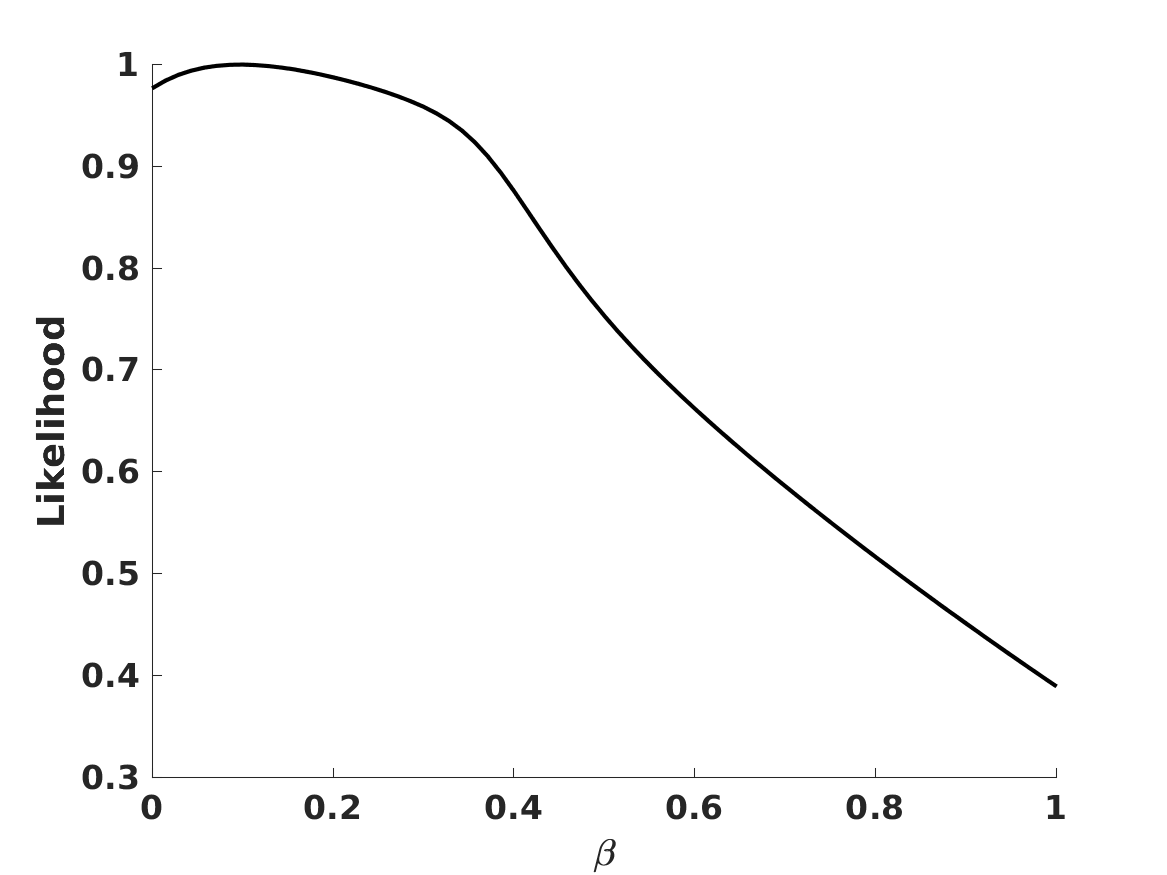}  
  \caption{One-dimensional (marginalized) posterior likelihoods for the model parameters corresponding to the analysis in Fig. \protect\ref{fig5}.\label{fig6}}
\end{figure*}

The results of this analysis are depicted in Figs. \ref{fig5} and \ref{fig6}.  The reduced chi-square at the four-dimensional best fit model is $\chi^2_\nu=0.98$, as in the previous subsection. In this case we find the same two sigma constraint for the matter density
\be
\Omega_m=0.28\pm0.04\,
\ee
at the two sigma ($95.4\%$) confidence level, together with the two sigma upper bounds
\be
\log_{10}{\lambda}<-5.36
\ee
\be
\log_{10}{(1+w_0)_C}<-1.21\,;
\ee
note that the latter stronger than the one in the previous subsection. Conversely, the constraint on $\beta$ is weakened, and as in the cosmology-only case it is unconstrained at the two sigma level, but one does get a one sigma upper bound
\be
\beta<0.67\,.
\ee

Again it is worthy of note that according to  Eq. \ref{vbetalambda} the relation between the equation of state and the dimensionless potential slope is $(1+w_0)_V=\lambda/3$ when $\beta=0$ and  $(1+w_0)_V=\lambda^2/9$ when $\beta=1$. In this case $\lambda$ is directly constrained by the Eotvos parameter limit, which also leads to a stronger constraint on $(1+w_0)_C$, and together, given the degeneracies between the parameters, these imply that constraints on $\beta$ are slightly relaxed.

\section{\label{outlook}Conclusions}

We have explored the low-redshift cosmological consequences of a new class of DBI models, which includes both the rolling tachyon field and the generalized Chaplygin gas models as particular limits. Each of these limiting models effectively provides a mechanism for a deviation of the value of the dark energy equation of state from its canonical (cosmological constant) value, which can be separately constrained. The main phenomenological difference between the two mechanisms is that the field dependence of the potential, which is characteristic of the rolling tachyon, also leads to variations of the fine-structure constant. The latter can be constrained through high-resolution astrophysical observations as well as local laboratory tests, and both of these provide key constraints.

Indeed, by using cosmological data alone (cf. Sect. \ref{part1}) one gets relatively mild constraints on both mechanisms, although these could be improved by including additional cosmological data sets, most notably from the cosmic microwave background---a task that is left for future work. On the other hand, including the astrophysical and local measurements improves constraints on the rolling tachyon sector by about seven orders of magnitude (cf. Sect. \ref{part2}), at least if this is parametrized as a deviation of the dark energy equation of state from its $\Lambda$CDM behaviour. Nevertheless, it should be noticed that these constraints have some dependence on the choice of explicit parametrization and of priors on the model parameters, as illustrated by our constraints on the potential slope (cf. Sect. \ref{part3}).

Our results confirm and strengthen the earlier analysis in \cite{Roll} indicating that in these models the potential is constrained to be extremely flat. From an observational point of view, the most interesting consequence of these results is that any rolling tachyon contribution to dark energy, described by the usual parameter $(1+w_0)_V$, is constrained to be so small as to be effectively indistinguishable from a cosmological constant, if one relies only on cosmological observations---and this is the case both for current facilities and for any foreseeable ones. However, such a component could in principle be identified through astrophysical tests of the stability of the fine-structure constant, for which the model predicts a specific redshift dependence. Our analysis therefore highlights both the intrinsic constraining power of astrophysical and local tests of the stability of $\alpha$, and their synergies with traditional cosmological observables, in probing fundamental cosmology and the mechanisms underlying the recent acceleration of the universe.


\begin{acknowledgments}
This work was financed by FEDER---Fundo Europeu de Desenvolvimento Regional funds through the COMPETE 2020---Operational Programme for Competitiveness and Internationalisation (POCI), and by Portuguese funds through FCT - Funda\c c\~ao para a Ci\^encia e a Tecnologia in the framework of the project POCI-01-0145-FEDER-028987 and PTDC/FIS-AST/28987/2017. 
\end{acknowledgments}

\bibliography{dbi}

\begin{thebibliography}{39}%
\makeatletter
\providecommand \@ifxundefined [1]{%
 \@ifx{#1\undefined}
}%
\providecommand \@ifnum [1]{%
 \ifnum #1\expandafter \@firstoftwo
 \else \expandafter \@secondoftwo
 \fi
}%
\providecommand \@ifx [1]{%
 \ifx #1\expandafter \@firstoftwo
 \else \expandafter \@secondoftwo
 \fi
}%
\providecommand \natexlab [1]{#1}%
\providecommand \enquote  [1]{``#1''}%
\providecommand \bibnamefont  [1]{#1}%
\providecommand \bibfnamefont [1]{#1}%
\providecommand \citenamefont [1]{#1}%
\providecommand \href@noop [0]{\@secondoftwo}%
\providecommand \href [0]{\begingroup \@sanitize@url \@href}%
\providecommand \@href[1]{\@@startlink{#1}\@@href}%
\providecommand \@@href[1]{\endgroup#1\@@endlink}%
\providecommand \@sanitize@url [0]{\catcode `\\12\catcode `\$12\catcode
  `\&12\catcode `\#12\catcode `\^12\catcode `\_12\catcode `\%12\relax}%
\providecommand \@@startlink[1]{}%
\providecommand \@@endlink[0]{}%
\providecommand \url  [0]{\begingroup\@sanitize@url \@url }%
\providecommand \@url [1]{\endgroup\@href {#1}{\urlprefix }}%
\providecommand \urlprefix  [0]{URL }%
\providecommand \Eprint [0]{\href }%
\providecommand \doibase [0]{http://dx.doi.org/}%
\providecommand \selectlanguage [0]{\@gobble}%
\providecommand \bibinfo  [0]{\@secondoftwo}%
\providecommand \bibfield  [0]{\@secondoftwo}%
\providecommand \translation [1]{[#1]}%
\providecommand \BibitemOpen [0]{}%
\providecommand \bibitemStop [0]{}%
\providecommand \bibitemNoStop [0]{.\EOS\space}%
\providecommand \EOS [0]{\spacefactor3000\relax}%
\providecommand \BibitemShut  [1]{\csname bibitem#1\endcsname}%
\let\auto@bib@innerbib\@empty
\bibitem [{\citenamefont {Copeland}\ \emph {et~al.}(2006)\citenamefont
  {Copeland}, \citenamefont {Sami},\ and\ \citenamefont
  {Tsujikawa}}]{Copeland}%
  \BibitemOpen
  \bibfield  {author} {\bibinfo {author} {\bibfnamefont {E.~J.}\ \bibnamefont
  {Copeland}}, \bibinfo {author} {\bibfnamefont {M.}~\bibnamefont {Sami}}, \
  and\ \bibinfo {author} {\bibfnamefont {S.}~\bibnamefont {Tsujikawa}},\ }\href
  {\doibase 10.1142/S021827180600942X} {\bibfield  {journal} {\bibinfo
  {journal} {Int. J. Mod. Phys. D}\ }\textbf {\bibinfo {volume} {15}},\
  \bibinfo {pages} {1753} (\bibinfo {year} {2006})},\ \Eprint
  {http://arxiv.org/abs/hep-th/0603057} {arXiv:hep-th/0603057} \BibitemShut
  {NoStop}%
\bibitem [{\citenamefont {Frieman}\ \emph {et~al.}(2008)\citenamefont
  {Frieman}, \citenamefont {Turner},\ and\ \citenamefont {Huterer}}]{Frieman}%
  \BibitemOpen
  \bibfield  {author} {\bibinfo {author} {\bibfnamefont {J.}~\bibnamefont
  {Frieman}}, \bibinfo {author} {\bibfnamefont {M.}~\bibnamefont {Turner}}, \
  and\ \bibinfo {author} {\bibfnamefont {D.}~\bibnamefont {Huterer}},\ }\href
  {\doibase 10.1146/annurev.astro.46.060407.145243} {\bibfield  {journal}
  {\bibinfo  {journal} {Ann. Rev. Astron. Astrophys.}\ }\textbf {\bibinfo
  {volume} {46}},\ \bibinfo {pages} {385} (\bibinfo {year} {2008})},\ \Eprint
  {http://arxiv.org/abs/0803.0982} {arXiv:0803.0982 [astro-ph]} \BibitemShut
  {NoStop}%
\bibitem [{\citenamefont {Joyce}\ \emph {et~al.}(2016)\citenamefont {Joyce},
  \citenamefont {Lombriser},\ and\ \citenamefont {Schmidt}}]{Joyce}%
  \BibitemOpen
  \bibfield  {author} {\bibinfo {author} {\bibfnamefont {A.}~\bibnamefont
  {Joyce}}, \bibinfo {author} {\bibfnamefont {L.}~\bibnamefont {Lombriser}}, \
  and\ \bibinfo {author} {\bibfnamefont {F.}~\bibnamefont {Schmidt}},\ }\href
  {\doibase 10.1146/annurev-nucl-102115-044553} {\bibfield  {journal} {\bibinfo
   {journal} {Ann. Rev. Nucl. Part. Sci.}\ }\textbf {\bibinfo {volume} {66}},\
  \bibinfo {pages} {95} (\bibinfo {year} {2016})},\ \Eprint
  {http://arxiv.org/abs/1601.06133} {arXiv:1601.06133 [astro-ph.CO]}
  \BibitemShut {NoStop}%
\bibitem [{\citenamefont {Huterer}\ and\ \citenamefont
  {Shafer}(2018)}]{Huterer}%
  \BibitemOpen
  \bibfield  {author} {\bibinfo {author} {\bibfnamefont {D.}~\bibnamefont
  {Huterer}}\ and\ \bibinfo {author} {\bibfnamefont {D.~L.}\ \bibnamefont
  {Shafer}},\ }\href {\doibase 10.1088/1361-6633/aa997e} {\bibfield  {journal}
  {\bibinfo  {journal} {Rept. Prog. Phys.}\ }\textbf {\bibinfo {volume} {81}},\
  \bibinfo {pages} {016901} (\bibinfo {year} {2018})},\ \Eprint
  {http://arxiv.org/abs/1709.01091} {arXiv:1709.01091 [astro-ph.CO]}
  \BibitemShut {NoStop}%
\bibitem [{\citenamefont {Martins}\ and\ \citenamefont
  {Moucherek}(2016)}]{Roll}%
  \BibitemOpen
  \bibfield  {author} {\bibinfo {author} {\bibfnamefont {C.~J. A.~P.}\
  \bibnamefont {Martins}}\ and\ \bibinfo {author} {\bibfnamefont {F.~M.~O.}\
  \bibnamefont {Moucherek}},\ }\href {\doibase 10.1103/PhysRevD.93.123524}
  {\bibfield  {journal} {\bibinfo  {journal} {Phys. Rev.}\ }\textbf {\bibinfo
  {volume} {D93}},\ \bibinfo {pages} {123524} (\bibinfo {year} {2016})},\
  \Eprint {http://arxiv.org/abs/1606.08380} {arXiv:1606.08380 [astro-ph.CO]}
  \BibitemShut {NoStop}%
\bibitem [{\citenamefont {Sen}(2002{\natexlab{a}})}]{Sen1}%
  \BibitemOpen
  \bibfield  {author} {\bibinfo {author} {\bibfnamefont {A.}~\bibnamefont
  {Sen}},\ }\href {\doibase 10.1088/1126-6708/2002/04/048} {\bibfield
  {journal} {\bibinfo  {journal} {JHEP}\ }\textbf {\bibinfo {volume} {04}},\
  \bibinfo {pages} {048} (\bibinfo {year} {2002}{\natexlab{a}})},\ \Eprint
  {http://arxiv.org/abs/hep-th/0203211} {arXiv:hep-th/0203211 [hep-th]}
  \BibitemShut {NoStop}%
\bibitem [{\citenamefont {Sen}(2002{\natexlab{b}})}]{Sen2}%
  \BibitemOpen
  \bibfield  {author} {\bibinfo {author} {\bibfnamefont {A.}~\bibnamefont
  {Sen}},\ }\href {\doibase 10.1088/1126-6708/2002/07/065} {\bibfield
  {journal} {\bibinfo  {journal} {JHEP}\ }\textbf {\bibinfo {volume} {07}},\
  \bibinfo {pages} {065} (\bibinfo {year} {2002}{\natexlab{b}})},\ \Eprint
  {http://arxiv.org/abs/hep-th/0203265} {arXiv:hep-th/0203265 [hep-th]}
  \BibitemShut {NoStop}%
\bibitem [{\citenamefont {Garousi}\ \emph {et~al.}(2005)\citenamefont
  {Garousi}, \citenamefont {Sami},\ and\ \citenamefont {Tsujikawa}}]{Garousi}%
  \BibitemOpen
  \bibfield  {author} {\bibinfo {author} {\bibfnamefont {M.~R.}\ \bibnamefont
  {Garousi}}, \bibinfo {author} {\bibfnamefont {M.}~\bibnamefont {Sami}}, \
  and\ \bibinfo {author} {\bibfnamefont {S.}~\bibnamefont {Tsujikawa}},\ }\href
  {\doibase 10.1103/PhysRevD.71.083005} {\bibfield  {journal} {\bibinfo
  {journal} {Phys. Rev.}\ }\textbf {\bibinfo {volume} {D71}},\ \bibinfo {pages}
  {083005} (\bibinfo {year} {2005})},\ \Eprint
  {http://arxiv.org/abs/hep-th/0412002} {arXiv:hep-th/0412002 [hep-th]}
  \BibitemShut {NoStop}%
\bibitem [{\citenamefont {Martins}(2017)}]{ROPP}%
  \BibitemOpen
  \bibfield  {author} {\bibinfo {author} {\bibfnamefont {C.~J. A.~P.}\
  \bibnamefont {Martins}},\ }\href {\doibase 10.1088/1361-6633/aa860e}
  {\bibfield  {journal} {\bibinfo  {journal} {Rep. Prog. Phys.}\ }\textbf
  {\bibinfo {volume} {80}},\ \bibinfo {pages} {126902} (\bibinfo {year}
  {2017})},\ \Eprint {http://arxiv.org/abs/1709.02923} {arXiv:1709.02923
  [astro-ph.CO]} \BibitemShut {NoStop}%
\bibitem [{\citenamefont {Bento}\ \emph {et~al.}(2002)\citenamefont {Bento},
  \citenamefont {Bertolami},\ and\ \citenamefont {Sen}}]{Bento}%
  \BibitemOpen
  \bibfield  {author} {\bibinfo {author} {\bibfnamefont {M.~C.}\ \bibnamefont
  {Bento}}, \bibinfo {author} {\bibfnamefont {O.}~\bibnamefont {Bertolami}}, \
  and\ \bibinfo {author} {\bibfnamefont {A.~A.}\ \bibnamefont {Sen}},\ }\href
  {\doibase 10.1103/PhysRevD.66.043507} {\bibfield  {journal} {\bibinfo
  {journal} {Phys. Rev.}\ }\textbf {\bibinfo {volume} {D66}},\ \bibinfo {pages}
  {043507} (\bibinfo {year} {2002})},\ \Eprint
  {http://arxiv.org/abs/gr-qc/0202064} {arXiv:gr-qc/0202064 [gr-qc]}
  \BibitemShut {NoStop}%
\bibitem [{\citenamefont {Be\c{c}a}(2008)}]{Beca}%
  \BibitemOpen
  \bibfield  {author} {\bibinfo {author} {\bibfnamefont {L.~M.}\ \bibnamefont
  {Be\c{c}a}},\ }\emph {\bibinfo {title} {{Unified Dark Energy Models}}},\
  \href@noop {} {Ph.D. thesis},\ \bibinfo  {school} {Porto U.} (\bibinfo {year}
  {2008})\BibitemShut {NoStop}%
\bibitem [{\citenamefont {Carroll}(1998)}]{Carroll}%
  \BibitemOpen
  \bibfield  {author} {\bibinfo {author} {\bibfnamefont {S.~M.}\ \bibnamefont
  {Carroll}},\ }\href {\doibase 10.1103/PhysRevLett.81.3067} {\bibfield
  {journal} {\bibinfo  {journal} {Phys.Rev.Lett.}\ }\textbf {\bibinfo {volume}
  {81}},\ \bibinfo {pages} {3067} (\bibinfo {year} {1998})},\ \Eprint
  {http://arxiv.org/abs/astro-ph/9806099} {arXiv:astro-ph/9806099 [astro-ph]}
  \BibitemShut {NoStop}%
\bibitem [{\citenamefont {Dvali}\ and\ \citenamefont
  {Zaldarriaga}(2002)}]{Dvali}%
  \BibitemOpen
  \bibfield  {author} {\bibinfo {author} {\bibfnamefont {G.}~\bibnamefont
  {Dvali}}\ and\ \bibinfo {author} {\bibfnamefont {M.}~\bibnamefont
  {Zaldarriaga}},\ }\href {\doibase 10.1103/PhysRevLett.88.091303} {\bibfield
  {journal} {\bibinfo  {journal} {Phys.Rev.Lett.}\ }\textbf {\bibinfo {volume}
  {88}},\ \bibinfo {pages} {091303} (\bibinfo {year} {2002})},\ \Eprint
  {http://arxiv.org/abs/hep-ph/0108217} {arXiv:hep-ph/0108217 [hep-ph]}
  \BibitemShut {NoStop}%
\bibitem [{\citenamefont {Chiba}\ and\ \citenamefont {Kohri}(2002)}]{Chiba}%
  \BibitemOpen
  \bibfield  {author} {\bibinfo {author} {\bibfnamefont {T.}~\bibnamefont
  {Chiba}}\ and\ \bibinfo {author} {\bibfnamefont {K.}~\bibnamefont {Kohri}},\
  }\href {\doibase 10.1143/PTP.107.631} {\bibfield  {journal} {\bibinfo
  {journal} {Prog.Theor.Phys.}\ }\textbf {\bibinfo {volume} {107}},\ \bibinfo
  {pages} {631} (\bibinfo {year} {2002})},\ \Eprint
  {http://arxiv.org/abs/hep-ph/0111086} {arXiv:hep-ph/0111086 [hep-ph]}
  \BibitemShut {NoStop}%
\bibitem [{\citenamefont {Damour}\ \emph {et~al.}(2002)\citenamefont {Damour},
  \citenamefont {Piazza},\ and\ \citenamefont {Veneziano}}]{Damour}%
  \BibitemOpen
  \bibfield  {author} {\bibinfo {author} {\bibfnamefont {T.}~\bibnamefont
  {Damour}}, \bibinfo {author} {\bibfnamefont {F.}~\bibnamefont {Piazza}}, \
  and\ \bibinfo {author} {\bibfnamefont {G.}~\bibnamefont {Veneziano}},\ }\href
  {\doibase 10.1103/PhysRevLett.89.081601} {\bibfield  {journal} {\bibinfo
  {journal} {Phys. Rev. Lett.}\ }\textbf {\bibinfo {volume} {89}},\ \bibinfo
  {pages} {081601} (\bibinfo {year} {2002})},\ \Eprint
  {http://arxiv.org/abs/gr-qc/0204094} {arXiv:gr-qc/0204094} \BibitemShut
  {NoStop}%
\bibitem [{\citenamefont {Nunes}\ and\ \citenamefont {Lidsey}(2004)}]{Lidsey}%
  \BibitemOpen
  \bibfield  {author} {\bibinfo {author} {\bibfnamefont {N.~J.}\ \bibnamefont
  {Nunes}}\ and\ \bibinfo {author} {\bibfnamefont {J.~E.}\ \bibnamefont
  {Lidsey}},\ }\href {\doibase 10.1103/PhysRevD.69.123511} {\bibfield
  {journal} {\bibinfo  {journal} {Phys.Rev.}\ }\textbf {\bibinfo {volume}
  {D69}},\ \bibinfo {pages} {123511} (\bibinfo {year} {2004})},\ \Eprint
  {http://arxiv.org/abs/astro-ph/0310882} {arXiv:astro-ph/0310882 [astro-ph]}
  \BibitemShut {NoStop}%
\bibitem [{\citenamefont {Amendola}\ \emph {et~al.}(2012)\citenamefont
  {Amendola}, \citenamefont {Leite}, \citenamefont {Martins}, \citenamefont
  {Nunes}, \citenamefont {Pedrosa} \emph {et~al.}}]{Amendola}%
  \BibitemOpen
  \bibfield  {author} {\bibinfo {author} {\bibfnamefont {L.}~\bibnamefont
  {Amendola}}, \bibinfo {author} {\bibfnamefont {A.~C.~O.}\ \bibnamefont
  {Leite}}, \bibinfo {author} {\bibfnamefont {C.~J. A.~P.}\ \bibnamefont
  {Martins}}, \bibinfo {author} {\bibfnamefont {N.}~\bibnamefont {Nunes}},
  \bibinfo {author} {\bibfnamefont {P.~O.~J.}\ \bibnamefont {Pedrosa}},  \emph
  {et~al.},\ }\href {\doibase 10.1103/PhysRevD.86.063515} {\bibfield  {journal}
  {\bibinfo  {journal} {Phys.Rev.}\ }\textbf {\bibinfo {volume} {D86}},\
  \bibinfo {pages} {063515} (\bibinfo {year} {2012})},\ \Eprint
  {http://arxiv.org/abs/1109.6793} {arXiv:1109.6793 [astro-ph.CO]} \BibitemShut
  {NoStop}%
\bibitem [{\citenamefont {Vielzeuf}\ and\ \citenamefont
  {Martins}(2013)}]{Pauline}%
  \BibitemOpen
  \bibfield  {author} {\bibinfo {author} {\bibfnamefont {P.~E.}\ \bibnamefont
  {Vielzeuf}}\ and\ \bibinfo {author} {\bibfnamefont {C.~J. A.~P.}\
  \bibnamefont {Martins}},\ }\href@noop {} {\  (\bibinfo {year} {2013})},\
  \Eprint {http://arxiv.org/abs/1309.7771} {arXiv:1309.7771 [astro-ph.CO]}
  \BibitemShut {NoStop}%
\bibitem [{\citenamefont {Gorini}\ \emph {et~al.}(2003)\citenamefont {Gorini},
  \citenamefont {Kamenshchik},\ and\ \citenamefont {Moschella}}]{Gorini}%
  \BibitemOpen
  \bibfield  {author} {\bibinfo {author} {\bibfnamefont {V.}~\bibnamefont
  {Gorini}}, \bibinfo {author} {\bibfnamefont {A.}~\bibnamefont {Kamenshchik}},
  \ and\ \bibinfo {author} {\bibfnamefont {U.}~\bibnamefont {Moschella}},\
  }\href {\doibase 10.1103/PhysRevD.67.063509} {\bibfield  {journal} {\bibinfo
  {journal} {Phys.\ Rev.\ D}\ }\textbf {\bibinfo {volume} {67}},\ \bibinfo
  {pages} {063509} (\bibinfo {year} {2003})},\ \Eprint
  {http://arxiv.org/abs/astro-ph/0209395} {arXiv:astro-ph/0209395} \BibitemShut
  {NoStop}%
\bibitem [{\citenamefont {Bean}\ and\ \citenamefont {Dore}(2003)}]{Bean}%
  \BibitemOpen
  \bibfield  {author} {\bibinfo {author} {\bibfnamefont {R.}~\bibnamefont
  {Bean}}\ and\ \bibinfo {author} {\bibfnamefont {O.}~\bibnamefont {Dore}},\
  }\href {\doibase 10.1103/PhysRevD.68.023515} {\bibfield  {journal} {\bibinfo
  {journal} {Phys.\ Rev.\ D}\ }\textbf {\bibinfo {volume} {68}},\ \bibinfo
  {pages} {023515} (\bibinfo {year} {2003})},\ \Eprint
  {http://arxiv.org/abs/astro-ph/0301308} {arXiv:astro-ph/0301308} \BibitemShut
  {NoStop}%
\bibitem [{\citenamefont {Amendola}\ \emph {et~al.}(2003)\citenamefont
  {Amendola}, \citenamefont {Finelli}, \citenamefont {Burigana},\ and\
  \citenamefont {Carturan}}]{Amendola2}%
  \BibitemOpen
  \bibfield  {author} {\bibinfo {author} {\bibfnamefont {L.}~\bibnamefont
  {Amendola}}, \bibinfo {author} {\bibfnamefont {F.}~\bibnamefont {Finelli}},
  \bibinfo {author} {\bibfnamefont {C.}~\bibnamefont {Burigana}}, \ and\
  \bibinfo {author} {\bibfnamefont {D.}~\bibnamefont {Carturan}},\ }\href
  {\doibase 10.1088/1475-7516/2003/07/005} {\bibfield  {journal} {\bibinfo
  {journal} {JCAP}\ }\textbf {\bibinfo {volume} {07}},\ \bibinfo {pages} {005}
  (\bibinfo {year} {2003})},\ \Eprint {http://arxiv.org/abs/astro-ph/0304325}
  {arXiv:astro-ph/0304325} \BibitemShut {NoStop}%
\bibitem [{\citenamefont {Sandvik}\ \emph {et~al.}(2004)\citenamefont
  {Sandvik}, \citenamefont {Tegmark}, \citenamefont {Zaldarriaga},\ and\
  \citenamefont {Waga}}]{Sandvik}%
  \BibitemOpen
  \bibfield  {author} {\bibinfo {author} {\bibfnamefont {H.}~\bibnamefont
  {Sandvik}}, \bibinfo {author} {\bibfnamefont {M.}~\bibnamefont {Tegmark}},
  \bibinfo {author} {\bibfnamefont {M.}~\bibnamefont {Zaldarriaga}}, \ and\
  \bibinfo {author} {\bibfnamefont {I.}~\bibnamefont {Waga}},\ }\href {\doibase
  10.1103/PhysRevD.69.123524} {\bibfield  {journal} {\bibinfo  {journal} {Phys.
  Rev. D}\ }\textbf {\bibinfo {volume} {69}},\ \bibinfo {pages} {123524}
  (\bibinfo {year} {2004})},\ \Eprint {http://arxiv.org/abs/astro-ph/0212114}
  {arXiv:astro-ph/0212114} \BibitemShut {NoStop}%
\bibitem [{\citenamefont {Park}\ \emph {et~al.}(2010)\citenamefont {Park},
  \citenamefont {Hwang}, \citenamefont {Park},\ and\ \citenamefont
  {Noh}}]{Park}%
  \BibitemOpen
  \bibfield  {author} {\bibinfo {author} {\bibfnamefont {C.-G.}\ \bibnamefont
  {Park}}, \bibinfo {author} {\bibfnamefont {J.-c.}\ \bibnamefont {Hwang}},
  \bibinfo {author} {\bibfnamefont {J.}~\bibnamefont {Park}}, \ and\ \bibinfo
  {author} {\bibfnamefont {H.}~\bibnamefont {Noh}},\ }\href {\doibase
  10.1103/PhysRevD.81.063532} {\bibfield  {journal} {\bibinfo  {journal} {Phys.
  Rev. D}\ }\textbf {\bibinfo {volume} {81}},\ \bibinfo {pages} {063532}
  (\bibinfo {year} {2010})},\ \Eprint {http://arxiv.org/abs/0910.4202}
  {arXiv:0910.4202 [astro-ph.CO]} \BibitemShut {NoStop}%
\bibitem [{\citenamefont {Marttens}\ \emph {et~al.}(2017)\citenamefont
  {Marttens}, \citenamefont {Casarini}, \citenamefont {Zimdahl}, \citenamefont
  {Hipolito-Ricaldi},\ and\ \citenamefont {Mota}}]{Marttens}%
  \BibitemOpen
  \bibfield  {author} {\bibinfo {author} {\bibfnamefont {R.~F.~v.}\
  \bibnamefont {Marttens}}, \bibinfo {author} {\bibfnamefont {L.}~\bibnamefont
  {Casarini}}, \bibinfo {author} {\bibfnamefont {W.}~\bibnamefont {Zimdahl}},
  \bibinfo {author} {\bibfnamefont {W.~S.}\ \bibnamefont {Hipolito-Ricaldi}}, \
  and\ \bibinfo {author} {\bibfnamefont {D.~F.}\ \bibnamefont {Mota}},\ }\href
  {\doibase 10.1016/j.dark.2017.02.001} {\bibfield  {journal} {\bibinfo
  {journal} {Phys. Dark Univ.}\ }\textbf {\bibinfo {volume} {15}},\ \bibinfo
  {pages} {114} (\bibinfo {year} {2017})},\ \Eprint
  {http://arxiv.org/abs/1702.00651} {arXiv:1702.00651 [astro-ph.CO]}
  \BibitemShut {NoStop}%
\bibitem [{\citenamefont {Feng}\ and\ \citenamefont {Li}(2014)}]{Feng}%
  \BibitemOpen
  \bibfield  {author} {\bibinfo {author} {\bibfnamefont {C.-J.}\ \bibnamefont
  {Feng}}\ and\ \bibinfo {author} {\bibfnamefont {X.-Z.}\ \bibnamefont {Li}},\
  }\href {\doibase 10.3969/j.issn.1000-5137.2014.04.015} {\ ,\ \bibinfo {pages}
  {432} (\bibinfo {year} {2014})},\ \Eprint {http://arxiv.org/abs/0909.5476}
  {arXiv:0909.5476 [astro-ph.CO]} \BibitemShut {NoStop}%
\bibitem [{\citenamefont {Wei}(2009)}]{Wei}%
  \BibitemOpen
  \bibfield  {author} {\bibinfo {author} {\bibfnamefont {H.}~\bibnamefont
  {Wei}},\ }\href {\doibase 10.1016/j.physletb.2009.10.086} {\bibfield
  {journal} {\bibinfo  {journal} {Phys. Lett.}\ }\textbf {\bibinfo {volume}
  {B682}},\ \bibinfo {pages} {98} (\bibinfo {year} {2009})},\ \Eprint
  {http://arxiv.org/abs/0907.2749} {arXiv:0907.2749 [gr-qc]} \BibitemShut
  {NoStop}%
\bibitem [{\citenamefont {Farajollahi}\ and\ \citenamefont
  {Tayebi}(2015)}]{Tayebi}%
  \BibitemOpen
  \bibfield  {author} {\bibinfo {author} {\bibfnamefont {H.}~\bibnamefont
  {Farajollahi}}\ and\ \bibinfo {author} {\bibfnamefont {F.}~\bibnamefont
  {Tayebi}},\ }\href {\doibase 10.1088/1674-4527/15/2/003} {\bibfield
  {journal} {\bibinfo  {journal} {Res. Astron. Astrophys.}\ }\textbf {\bibinfo
  {volume} {15}},\ \bibinfo {pages} {191} (\bibinfo {year} {2015})}\BibitemShut
  {NoStop}%
\bibitem [{\citenamefont {Steinhardt}\ \emph {et~al.}(2020)\citenamefont
  {Steinhardt}, \citenamefont {Sneppen},\ and\ \citenamefont
  {Sen}}]{Steinhardt}%
  \BibitemOpen
  \bibfield  {author} {\bibinfo {author} {\bibfnamefont {C.~L.}\ \bibnamefont
  {Steinhardt}}, \bibinfo {author} {\bibfnamefont {A.}~\bibnamefont {Sneppen}},
  \ and\ \bibinfo {author} {\bibfnamefont {B.}~\bibnamefont {Sen}},\ }\href
  {\doibase 10.3847/1538-4357/abb140} {\bibfield  {journal} {\bibinfo
  {journal} {Astrophys. J.}\ }\textbf {\bibinfo {volume} {902}},\ \bibinfo
  {pages} {14} (\bibinfo {year} {2020})},\ \Eprint
  {http://arxiv.org/abs/2005.07707} {arXiv:2005.07707 [astro-ph.CO]}
  \BibitemShut {NoStop}%
\bibitem [{\citenamefont {Riess}\ \emph {et~al.}(2018)\citenamefont {Riess}
  \emph {et~al.}}]{Riess}%
  \BibitemOpen
  \bibfield  {author} {\bibinfo {author} {\bibfnamefont {A.~G.}\ \bibnamefont
  {Riess}} \emph {et~al.},\ }\href {\doibase 10.3847/1538-4357/aaa5a9}
  {\bibfield  {journal} {\bibinfo  {journal} {Astrophys. J.}\ }\textbf
  {\bibinfo {volume} {853}},\ \bibinfo {pages} {126} (\bibinfo {year}
  {2018})},\ \Eprint {http://arxiv.org/abs/1710.00844} {arXiv:1710.00844
  [astro-ph.CO]} \BibitemShut {NoStop}%
\bibitem [{\citenamefont {Farooq}\ \emph {et~al.}(2017)\citenamefont {Farooq},
  \citenamefont {Madiyar}, \citenamefont {Crandall},\ and\ \citenamefont
  {Ratra}}]{Farooq}%
  \BibitemOpen
  \bibfield  {author} {\bibinfo {author} {\bibfnamefont {O.}~\bibnamefont
  {Farooq}}, \bibinfo {author} {\bibfnamefont {F.~R.}\ \bibnamefont {Madiyar}},
  \bibinfo {author} {\bibfnamefont {S.}~\bibnamefont {Crandall}}, \ and\
  \bibinfo {author} {\bibfnamefont {B.}~\bibnamefont {Ratra}},\ }\href
  {\doibase 10.3847/1538-4357/835/1/26} {\bibfield  {journal} {\bibinfo
  {journal} {Astrophys. J.}\ }\textbf {\bibinfo {volume} {835}},\ \bibinfo
  {pages} {26} (\bibinfo {year} {2017})},\ \Eprint
  {http://arxiv.org/abs/1607.03537} {arXiv:1607.03537 [astro-ph.CO]}
  \BibitemShut {NoStop}%
\bibitem [{\citenamefont {Anagnostopoulos}\ and\ \citenamefont
  {Basilakos}(2018)}]{Anagnostopoulos}%
  \BibitemOpen
  \bibfield  {author} {\bibinfo {author} {\bibfnamefont {F.~K.}\ \bibnamefont
  {Anagnostopoulos}}\ and\ \bibinfo {author} {\bibfnamefont {S.}~\bibnamefont
  {Basilakos}},\ }\href {\doibase 10.1103/PhysRevD.97.063503} {\bibfield
  {journal} {\bibinfo  {journal} {Phys. Rev. D}\ }\textbf {\bibinfo {volume}
  {97}},\ \bibinfo {pages} {063503} (\bibinfo {year} {2018})},\ \Eprint
  {http://arxiv.org/abs/1709.02356} {arXiv:1709.02356 [astro-ph.CO]}
  \BibitemShut {NoStop}%
\bibitem [{\citenamefont {Webb}\ \emph {et~al.}(2011)\citenamefont {Webb},
  \citenamefont {King}, \citenamefont {Murphy}, \citenamefont {Flambaum},
  \citenamefont {Carswell} \emph {et~al.}}]{Webb}%
  \BibitemOpen
  \bibfield  {author} {\bibinfo {author} {\bibfnamefont {J.}~\bibnamefont
  {Webb}}, \bibinfo {author} {\bibfnamefont {J.}~\bibnamefont {King}}, \bibinfo
  {author} {\bibfnamefont {M.}~\bibnamefont {Murphy}}, \bibinfo {author}
  {\bibfnamefont {V.}~\bibnamefont {Flambaum}}, \bibinfo {author}
  {\bibfnamefont {R.}~\bibnamefont {Carswell}},  \emph {et~al.},\ }\href
  {\doibase 10.1103/PhysRevLett.107.191101} {\bibfield  {journal} {\bibinfo
  {journal} {Phys.Rev.Lett.}\ }\textbf {\bibinfo {volume} {107}},\ \bibinfo
  {pages} {191101} (\bibinfo {year} {2011})},\ \Eprint
  {http://arxiv.org/abs/1008.3907} {arXiv:1008.3907 [astro-ph.CO]} \BibitemShut
  {NoStop}%
\bibitem [{\citenamefont {Murphy}\ and\ \citenamefont
  {Cooksey}(2017)}]{Cooksey}%
  \BibitemOpen
  \bibfield  {author} {\bibinfo {author} {\bibfnamefont {M.~T.}\ \bibnamefont
  {Murphy}}\ and\ \bibinfo {author} {\bibfnamefont {K.~L.}\ \bibnamefont
  {Cooksey}},\ }\href {\doibase 10.1093/mnras/stx1949} {\bibfield  {journal}
  {\bibinfo  {journal} {Mon. Not. Roy. Astron. Soc.}\ }\textbf {\bibinfo
  {volume} {471}},\ \bibinfo {pages} {4930} (\bibinfo {year} {2017})},\ \Eprint
  {http://arxiv.org/abs/1708.00014} {arXiv:1708.00014 [astro-ph.CO]}
  \BibitemShut {NoStop}%
\bibitem [{\citenamefont {Welsh}\ \emph {et~al.}(2020)\citenamefont {Welsh},
  \citenamefont {Cooke}, \citenamefont {Fumagalli},\ and\ \citenamefont
  {Pettini}}]{Welsh}%
  \BibitemOpen
  \bibfield  {author} {\bibinfo {author} {\bibfnamefont {L.}~\bibnamefont
  {Welsh}}, \bibinfo {author} {\bibfnamefont {R.}~\bibnamefont {Cooke}},
  \bibinfo {author} {\bibfnamefont {M.}~\bibnamefont {Fumagalli}}, \ and\
  \bibinfo {author} {\bibfnamefont {M.}~\bibnamefont {Pettini}},\ }\href
  {\doibase 10.1093/mnras/staa807} {\bibfield  {journal} {\bibinfo  {journal}
  {Monthly Notices of the Royal Astronomical Society}\ }\textbf {\bibinfo
  {volume} {494}},\ \bibinfo {pages} {1411–1423} (\bibinfo {year}
  {2020})}\BibitemShut {NoStop}%
\bibitem [{\citenamefont {Milakovi\'c}\ \emph {et~al.}(2020)\citenamefont
  {Milakovi\'c}, \citenamefont {Lee}, \citenamefont {Carswell}, \citenamefont
  {Webb}, \citenamefont {Molaro},\ and\ \citenamefont {Pasquini}}]{Milakovic}%
  \BibitemOpen
  \bibfield  {author} {\bibinfo {author} {\bibfnamefont {D.}~\bibnamefont
  {Milakovi\'c}}, \bibinfo {author} {\bibfnamefont {C.-C.}\ \bibnamefont
  {Lee}}, \bibinfo {author} {\bibfnamefont {R.~F.}\ \bibnamefont {Carswell}},
  \bibinfo {author} {\bibfnamefont {J.~K.}\ \bibnamefont {Webb}}, \bibinfo
  {author} {\bibfnamefont {P.}~\bibnamefont {Molaro}}, \ and\ \bibinfo {author}
  {\bibfnamefont {L.}~\bibnamefont {Pasquini}},\ }\href {\doibase
  10.1093/mnras/staa3217} {\  (\bibinfo {year} {2020}),\
  10.1093/mnras/staa3217},\ \Eprint {http://arxiv.org/abs/2008.10619}
  {arXiv:2008.10619 [astro-ph.CO]} \BibitemShut {NoStop}%
\bibitem [{\citenamefont {Martins}\ and\ \citenamefont {Vila
  Mi\~nana}(2019)}]{Meritxell}%
  \BibitemOpen
  \bibfield  {author} {\bibinfo {author} {\bibfnamefont {C.~J. A.~P.}\
  \bibnamefont {Martins}}\ and\ \bibinfo {author} {\bibfnamefont
  {M.}~\bibnamefont {Vila Mi\~nana}},\ }\href {\doibase
  10.1016/j.dark.2019.100301} {\bibfield  {journal} {\bibinfo  {journal} {Phys.
  Dark Univ.}\ }\textbf {\bibinfo {volume} {25}},\ \bibinfo {pages} {100301}
  (\bibinfo {year} {2019})},\ \Eprint {http://arxiv.org/abs/1904.07896}
  {arXiv:1904.07896 [astro-ph.CO]} \BibitemShut {NoStop}%
\bibitem [{\citenamefont {Petrov}\ \emph {et~al.}(2006)\citenamefont {Petrov},
  \citenamefont {Nazarov}, \citenamefont {Onegin}, \citenamefont {Petrov},\
  and\ \citenamefont {Sakhnovsky}}]{Oklo}%
  \BibitemOpen
  \bibfield  {author} {\bibinfo {author} {\bibfnamefont {{\relax Yu}.~V.}\
  \bibnamefont {Petrov}}, \bibinfo {author} {\bibfnamefont {A.~I.}\
  \bibnamefont {Nazarov}}, \bibinfo {author} {\bibfnamefont {M.~S.}\
  \bibnamefont {Onegin}}, \bibinfo {author} {\bibfnamefont {V.~{\relax Yu}.}\
  \bibnamefont {Petrov}}, \ and\ \bibinfo {author} {\bibfnamefont {E.~G.}\
  \bibnamefont {Sakhnovsky}},\ }\href {\doibase 10.1103/PhysRevC.74.064610}
  {\bibfield  {journal} {\bibinfo  {journal} {Phys. Rev.}\ }\textbf {\bibinfo
  {volume} {C74}},\ \bibinfo {pages} {064610} (\bibinfo {year} {2006})},\
  \Eprint {http://arxiv.org/abs/hep-ph/0506186} {arXiv:hep-ph/0506186 [hep-ph]}
  \BibitemShut {NoStop}%
\bibitem [{\citenamefont {Lange}\ \emph {et~al.}(2020)\citenamefont {Lange},
  \citenamefont {Huntemann}, \citenamefont {Rahm}, \citenamefont {Sanner},
  \citenamefont {Shao}, \citenamefont {Lipphardt}, \citenamefont {Tamm},
  \citenamefont {Weyers},\ and\ \citenamefont {Peik}}]{Lange}%
  \BibitemOpen
  \bibfield  {author} {\bibinfo {author} {\bibfnamefont {R.}~\bibnamefont
  {Lange}}, \bibinfo {author} {\bibfnamefont {N.}~\bibnamefont {Huntemann}},
  \bibinfo {author} {\bibfnamefont {J.}~\bibnamefont {Rahm}}, \bibinfo {author}
  {\bibfnamefont {C.}~\bibnamefont {Sanner}}, \bibinfo {author} {\bibfnamefont
  {H.}~\bibnamefont {Shao}}, \bibinfo {author} {\bibfnamefont {B.}~\bibnamefont
  {Lipphardt}}, \bibinfo {author} {\bibfnamefont {C.}~\bibnamefont {Tamm}},
  \bibinfo {author} {\bibfnamefont {S.}~\bibnamefont {Weyers}}, \ and\ \bibinfo
  {author} {\bibfnamefont {E.}~\bibnamefont {Peik}},\ }\href@noop {} {\
  (\bibinfo {year} {2020})},\ \Eprint {http://arxiv.org/abs/2010.06620}
  {arXiv:2010.06620 [physics.atom-ph]} \BibitemShut {NoStop}%
\bibitem [{\citenamefont {Touboul}\ \emph {et~al.}(2019)\citenamefont {Touboul}
  \emph {et~al.}}]{Touboul}%
  \BibitemOpen
  \bibfield  {author} {\bibinfo {author} {\bibfnamefont {P.}~\bibnamefont
  {Touboul}} \emph {et~al.} (\bibinfo {collaboration} {MICROSCOPE}),\ }\href
  {\doibase 10.1088/1361-6382/ab4707} {\bibfield  {journal} {\bibinfo
  {journal} {Class. Quant. Grav.}\ }\textbf {\bibinfo {volume} {36}},\ \bibinfo
  {pages} {225006} (\bibinfo {year} {2019})},\ \Eprint
  {http://arxiv.org/abs/1909.10598} {arXiv:1909.10598 [gr-qc]} \BibitemShut
  {NoStop}%
\end{thebibliography}%
\end{document}